\documentclass[amsmath,amssymb,aps,superscriptaddress,nofootinbib,showkeys,preprint,A4paper]{revtex4-2}

\usepackage[english]{babel}
\usepackage[utf8]{inputenc}
\usepackage{xr-hyper}

\usepackage[colorlinks]{hyperref}
\usepackage[T1]{fontenc}
\usepackage[dvips]{graphicx}
\usepackage{amsmath}
\usepackage{amsfonts}
\usepackage[dvipsnames]{xcolor}
\usepackage[caption=false, justification=justified]{subfig}
\usepackage{url}
\usepackage{ulem}
\usepackage{braket}

\makeatletter
\newcommand*{\addFileDependency}[1]{
  \typeout{(#1)}
  \@addtofilelist{#1}
  \IfFileExists{#1}{}{\typeout{No file #1.}}
}
\makeatother

\DeclareMathOperator{\Tr}{Tr}

\newcommand{\micro}{µ} 

\newcommand{\ang}{Å} 
\newcommand{\reb}[1]{\textcolor{blue}{#1}} 
\newcommand{\etal}{~\textit{et~al.}~}
\newcommand{\etc}{~\textit{etc.}}
\newcommand{\ie}{\textit{i.e.}}
\newcommand{\vs}{\textit{vs.}~}
\newcommand{\eg}{\textit{e.g.}}
\newcommand{\meths}[0]{\hyperref[meth]{Methods}}

\newcommand*{\myexternaldocument}[1]{%
    \externaldocument{#1}%
    \addFileDependency{#1.tex}%
    \addFileDependency{#1.aux}%
}

\myexternaldocument{supp_mat}

\hypersetup{filecolor=blue}
\hypersetup{citecolor=blue}
\hypersetup{urlcolor=blue}
\hypersetup{linkcolor=blue}

\begin{document}

\title{Spin-momentum locking and ultrafast spin-charge conversion in ultrathin epitaxial Bi$_{1-x}$Sb$_x$ topological insulator.}
\date{\today}
\author{E.~Rongione}
\affiliation{Unité Mixte de Physique, CNRS, Thales, Université Paris-Saclay, F-91767 Palaiseau, France}
\affiliation{Laboratoire de Physique de l’Ecole Normale Supérieure, ENS, Université PSL, CNRS, Sorbonne Université,
Université Paris Cité, F-75005 Paris, France}
\author{L.~Baringthon}
\affiliation{Unité Mixte de Physique, CNRS, Thales, Université Paris-Saclay, F-91767 Palaiseau, France}
\affiliation{Synchrotron SOLEIL, L’Orme des Merisiers, Départementale 128, F-91190 Saint-Aubin, France}
\affiliation{Université Paris-Saclay, CNRS, Centre de Nanosciences et de Nanotechnologies, F-91120 Palaiseau, France}
\author{D.~She}
\affiliation{Unité Mixte de Physique, CNRS, Thales, Université Paris-Saclay, F-91767 Palaiseau, France}
\affiliation{Synchrotron SOLEIL, L’Orme des Merisiers, Départementale 128, F-91190 Saint-Aubin, France}
\affiliation{Université Paris-Saclay, CNRS, Centre de Nanosciences et de Nanotechnologies, F-91120 Palaiseau, France}
\author{G.~Patriarche}
\affiliation{Université Paris-Saclay, CNRS, Centre de Nanosciences et de Nanotechnologies, F-91120 Palaiseau, France}
\author{R.~Lebrun}
\affiliation{Unité Mixte de Physique, CNRS, Thales, Université Paris-Saclay, F-91767 Palaiseau, France}
\author{A.~Lema\^{\i}tre}
\affiliation{Université Paris-Saclay, CNRS, Centre de Nanosciences et de Nanotechnologies, F-91120 Palaiseau, France}
\author{M. Morassi}
\affiliation{Université Paris-Saclay, CNRS, Centre de Nanosciences et de Nanotechnologies, F-91120 Palaiseau, France}
\author{N.~Reyren}
\affiliation{Unité Mixte de Physique, CNRS, Thales, Université Paris-Saclay, F-91767 Palaiseau, France}
\author{M.~Mičica}
\affiliation{Laboratoire de Physique de l’Ecole Normale Supérieure, ENS, Université PSL, CNRS, Sorbonne Université,
Université Paris Cité, F-75005 Paris, France}
\author{J.~Mangeney}
\affiliation{Laboratoire de Physique de l’Ecole Normale Supérieure, ENS, Université PSL, CNRS, Sorbonne Université,
Université Paris Cité, F-75005 Paris, France}
\author{J.~Tignon}
\affiliation{Laboratoire de Physique de l’Ecole Normale Supérieure, ENS, Université PSL, CNRS, Sorbonne Université,
Université Paris Cité, F-75005 Paris, France}
\author{F. Bertran}
\affiliation{Synchrotron SOLEIL, L’Orme des Merisiers, Départementale 128, F-91190 Saint-Aubin, France}
\author{S.~Dhillon}
\affiliation{Laboratoire de Physique de l’Ecole Normale Supérieure, ENS, Université PSL, CNRS, Sorbonne Université,
Université Paris Cité, F-75005 Paris, France}
\affiliation{\normalfont Corresponding authors:~\href{mailto:sukhdeep.dhillon@phys.ens.fr}{sukhdeep.dhillon@phys.ens.fr},~\href{mailto:henri.jaffres@cnrs-thales.fr}{henri.jaffres@cnrs-thales.fr},~\href{mailto:jeanmarie.george@cnrs-thales.fr}{jeanmarie.george@cnrs-thales.fr}}
\author{P.~Le Fèvre}
\affiliation{Synchrotron SOLEIL, L’Orme des Merisiers, Départementale 128, F-91190 Saint-Aubin, France}
\author{H.~Jaffrès}
\affiliation{Unité Mixte de Physique, CNRS, Thales, Université Paris-Saclay, F-91767 Palaiseau, France}
\affiliation{\normalfont Corresponding authors:~\href{mailto:sukhdeep.dhillon@phys.ens.fr}{sukhdeep.dhillon@phys.ens.fr},~\href{mailto:henri.jaffres@cnrs-thales.fr}{henri.jaffres@cnrs-thales.fr},~\href{mailto:jeanmarie.george@cnrs-thales.fr}{jeanmarie.george@cnrs-thales.fr}}
\author{J.-M.~George}
\affiliation{Unité Mixte de Physique, CNRS, Thales, Université Paris-Saclay, F-91767 Palaiseau, France}
\affiliation{\normalfont Corresponding authors:~\href{mailto:sukhdeep.dhillon@phys.ens.fr}{sukhdeep.dhillon@phys.ens.fr},~\href{mailto:henri.jaffres@cnrs-thales.fr}{henri.jaffres@cnrs-thales.fr},~\href{mailto:jeanmarie.george@cnrs-thales.fr}{jeanmarie.george@cnrs-thales.fr}}


\begin{abstract}
\textbf{The helicity of 3D topological insulator surface states has drawn significant attention in spintronics owing to spin-momentum locking where the carriers' spin is oriented perpendicular to their momentum. This property can provide an efficient method to convert charge currents into spin currents, and vice-versa, through the Rashba-Edelstein effect. However, experimental signatures of these surface states to the spin-charge conversion are extremely difficult to disentangle from bulk state contributions. Here, we combine spin- and angle-resolved photo-emission spectroscopy, and time-resolved THz emission spectroscopy to categorically demonstrate that spin-charge conversion arises mainly from the surface state in Bi$_{1-x}$Sb$_x$ ultrathin films, down to few nanometers where confinement effects emerge. We correlate this large conversion efficiency, typically at the level of the bulk spin Hall effect from heavy metals, to the complex Fermi surface obtained from theoretical calculations of the inverse Rashba-Edelstein response. 
Both surface state robustness and sizeable conversion efficiency in epitaxial Bi$_{1-x}$Sb$_x$ thin films bring new perspectives for ultra-low power magnetic random-access memories and broadband THz generation.}
\end{abstract}


\maketitle

\newpage

\begin{sloppypar}

\section*{Introduction}


The discovery of metallic quantum states at the surface of 3D topological insulators (TIs)~\cite{hasan_colloquium_2010,ando_topological_2013,qi_topological_2011} has opened exciting new functionalities in spintronics owing to their topological protection and spin-momentum locking (SML) properties~\cite{shiomi_spin-electricity_2014,ando_spin_2017}. Indeed, the combination of band inversion and time reversal symmetry (TRS) results in a peculiar spin texture in momentum space. Injecting a current in these states results therefore in an out of equilibrium spin density (also called spin-accumulation) along the transverse direction~\cite{shiomi_spin-electricity_2014,ando_spin_2017,han_quantum_2018}. This is the Rashba-Edelstein effect (REE)~\cite{edelstein_spin_1990} which can be used to exert a spin-orbit torque (SOT) onto the magnetization of an adjacent ferromagnet (FM)~\cite{Mellnik2014}. The reciprocal phenomenon, by which a spin density produces an in-plane transverse charge current, is called the Inverse Rashba-Edelstein effect (IREE)~\cite{zhang_conversion_2016,han_quantum_2018}. 

Importantly, the resulting spin-charge conversion (SCC) efficiencies in topological surface states (TSS)  combining strong spin-orbit coupling (SOC) and SML is expected to be at least one order of magnitude larger compared to the spin Hall effect (SHE) of 5\textit{d} heavy metals~\cite{miron_perpendicular_2011,liu_current-induced_2012,liu_spin-torque_2012}. SCC has been demonstrated in a range of Bi-based TI coumpounds, including bismuth selenide Bi$_2$Se$_3$, bismuth telluride Bi$_2$Te$_3$, Bi$_2$(Se,Te)$_3$~\cite{Kondou2016} or Bi$_{1-x}$Sb$_x$~(BiSb)~\cite{khang_conductive_2018}. To benefit fully from IREE, the charge currents should be confined in the surface states and any current flowing through the bulk states should be avoided. The prerequisites are hence \textit{i}) a sizeable bandgap typically larger than 0.2 to 0.3~eV, and \textit{ii}) a perfect control of the Fermi level position, usually achieved by stoichiometry and/or strain engineering.  In this respect, Bi$_{1-x}$Sb$_{x}$ alloys, although displaying clear topological surface states~\cite{teo2008,hsieh_topological_2008,benia_surface_2015}, have been mostly neglected for spintronic applications as a result of their modest bulk bandgap (about 40~meV for $x=0.07$) and relatively complex band structure. However, quantization effects in ultrathin films have shown to lead to much larger gap~\cite{baringthon_topological_2022,ito2020} while  retaining their band inversion near the $\overline{\text{M}}$ point in the $x=0.07-0.3$ composition range~\cite{lenoir_chapter_2001}, unlike pure Bi~\cite{hirahara2006,koroteev2008,aguilera_z_2021}. BiSb therefore has considerable potential as candidate for spintronics applications as well as for recently engineered efficient spintronic THz emitters~\cite{Seifert2016,Wang2018,Tong2021,Chen2021,sharma2021,Park21,park_topological_2022,Rongione2022}.


In this letter, we report on our detailed investigation of the surface state SML properties of ultrathin (1\,1\,1)-oriented Bi$_{1-x}$Sb$_x$ epitaxial films. They exhibit a topological phase as recently confirmed by our angular-resolved photo-emission spectroscopy (ARPES) measurements~\cite{baringthon_topological_2022}. Here, we focus on spin-resolved ARPES (SARPES) performed on ultrathin BiSb films and extract the in-plane spin texture for the different electron and hole pockets characterizing the complex BiSb Fermi surface. Moreover, the SCC mediated by the BiSb surface states is probed at the sub-picosecond timescale using an adjacent metallic Co layer acting as a spin injector. Unprecedentedly large SCC is measured with efficiencies beyond the level of carefully optimized Co/Pt systems. Our results also indicate that surface state related IREE is the mechanism responsible for SCC mechanism. Tight-binding (TB) calculation and linear response theory account for our findings.



\section{Spin-resolved ARPES}

\begin{figure*}[!htp]
  \begin{center}
      \includegraphics[width=\textwidth]{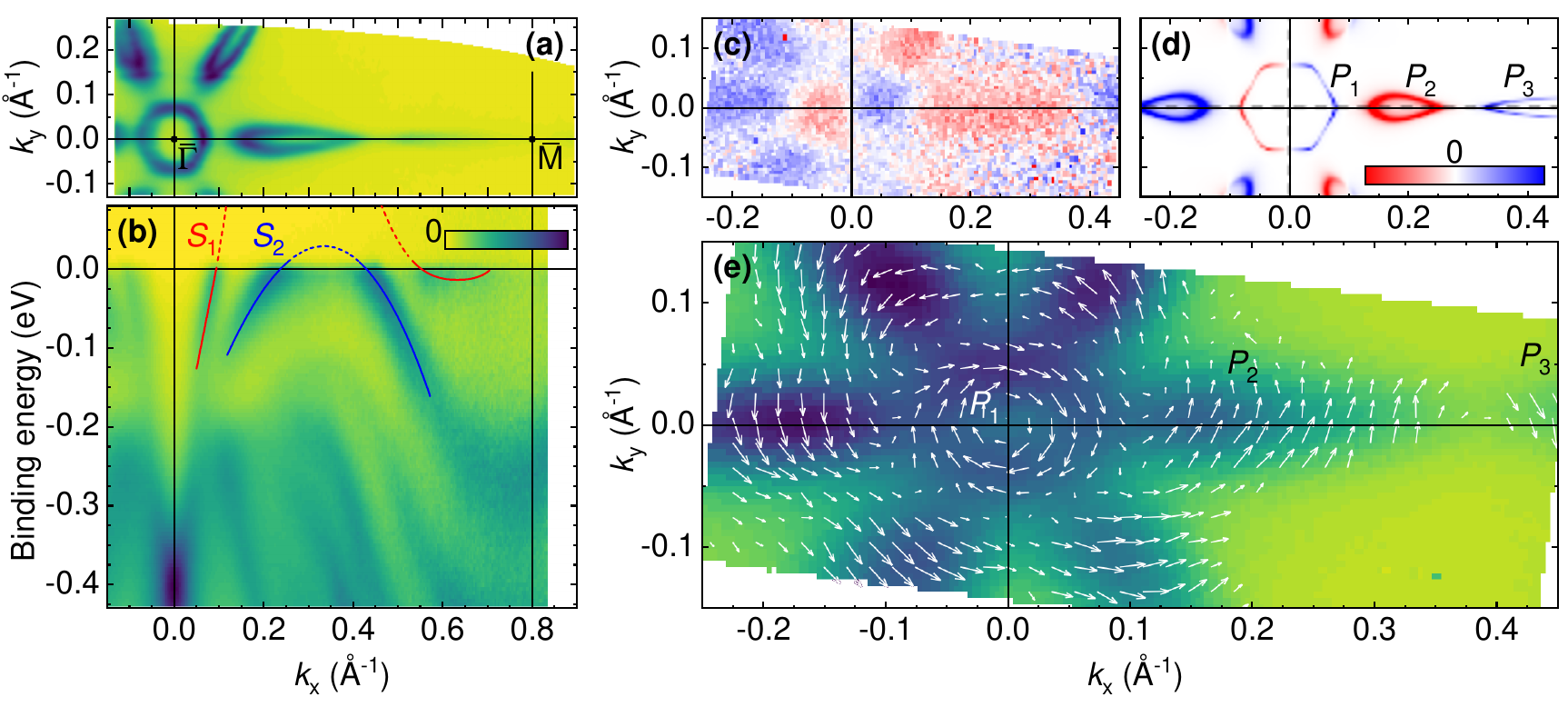}
       \caption{\textbf{Spin-resolved surface states of a 5-nm thick Bi$_{0.85}$Sb$_{0.15}$ film grown on Si(1\,1\,1).} 
       (a) High-resolution ARPES map at the Fermi energy (integrated over 20~meV). (b) ARPES energy dispersion along the $\overline{\Gamma}\overline{\text{M}}$ direction ($k_x$ direction). As a guide to the eye, red and blue lines underline the $S_1$ and $S_2$ surface states. All ARPES measurements were performed at 20 K. (c) $\sigma_y$ polarization DOS measured at the Fermi level, and (d) corresponding TB modelling of the s-DOS projected on the first BL. The experiment is performed at room temperature, while the calculations are at $T=0$. The color bar indicates the spin polarization $\sigma_y$ between -1 and +1 in (d). The color scale in (c) is proportional to the polarization with scale extrema of $P\cdot S=\pm0.1$. (e) Color map representing the measured ARPES intensity (arb. units, proportional to the DOS at the Fermi level) close to Fermi energy integrated on 25~meV with arrows representing the measured spin polarization direction and amplitude at room temperature. The different electron and hole pockets labeled $P_1$, $P_2$ and $P_3$ in (d) are easily identified in the experimental measurements.}
    \label{figSARPES}
  \end{center}
\end{figure*}

SARPES is the method of choice to probe the spin-textured Fermi contour of TI surfaces. Ultrathin epitaxial Bi$_{1-x}$Sb$_x$ films with $x=0.07$, 0.1, 0.15, 0.21, 0.3, 0.4 and thicknesses down to 2.5~nm were grown by molecular beam epitaxy (MBE) on clean $7\times7$ reconstructed Si(1\,1\,1) surfaces. They all exhibit a non-trivial topological phase. Details of the growth~\cite{baringthon_elaboration_2022} are quickly recalled in the \meths~section where scanning transmission electron microscopy (STEM) and energy-dispersive X-ray spectroscopy (EDX) characterizations are also discussed. 
We focus first on the 5~nm thick Bi$_{0.85}$Sb$_{0.15}$ sample. Fig.~\ref{figSARPES}\reb{a} shows its experimental Fermi surface within the 2D surface Brillouin zone, as measured by ARPES. It is composed of three pockets labelled $P_1$ (the hexagonal electron pocket surrounding $\overline{\Gamma}$), and two elongated $P_2$ hole and $P_3$ electron pockets along each of the six equivalent $\overline{\Gamma}\overline{\text{M}}$ directions (one has been chosen as the $k_x$ axis). Fig.~\ref{figSARPES}\reb{b} displays the  energy dispersion along the $\overline{\Gamma}\overline{\text{M}}$ direction. The signature of two surface $S_1$ and $S_2$ states~\cite{benia_surface_2015,baringthon_topological_2022} are visible in the bandgap.
The valence band state energy dispersion is also visible, from a series of confined states with energy splitting increasing with reducing thickness~\cite{baringthon_topological_2022}. The corresponding experimental SARPES polarization map for the $\sigma_y$ spin-polarization component is given in Fig.~\ref{figSARPES}\reb{c}. For this experiment, the incident photon energy is $20$~eV (5~meV resolution).

Owing to the electron analyzer movable entrance optics, the $\sigma_x$ and $\sigma_y$ spin polarizations were measured on a large part of the surface first Brillouin zone (see \meths). On Fig.~\ref{figSARPES}\reb{e}, the spin polarization is represented as a vector field. For clarity, the measured polarization is only displayed for positions where the DOS is the largest. The norm is roughly proportional to the spin-resolved DOS (s-DOS). The vector field is superimposed over a color map yielding the sum of all the signals measured by the spin detector, which is proportional to the DOS at the Fermi level. For a clearer representation, the vector fields were averaged over $0.02 \times 0.02$~\ang$^{-2}$ areas, corresponding to twenty data points. The experimental data reveals the helical spin texture of the inner $P_1$ Fermi contour, the opposite spin polarization of the $P_2$ hole pocket, and finally, the same spin chirality for weaker $P_3$ pocket. We recover from experiments the symmetry property imposing the orthogonality between the spin direction and the vertical symmetry planes $\sigma_V^{\overline{\Gamma\text{M}}}$ containing the $\overline{\Gamma}\overline{\text{M}}$ lines. Such property remains partly true concerning the in-plane spin components for the $\overline{\Gamma}\overline{\text{K}}$ directions even if the corresponding vertical planes $\sigma_V^{\overline{\Gamma\text{K}}}$ do not represent perfect symmetry operators. The lack of symmetry for $\sigma_V^{\overline{\Gamma\text{K}}}$ leads to the warping term responsible for the appearance of a $\sigma_z$ component~\cite{ando_topological_2013}, not presently discussed.

We now compare the $S_1$ and $S_2$ surface state spin-texture with a TB model. Calculations are implemented by considering  
relaxed bulk lattice parameters following the Vegard’s law (refer to Ref.~\cite{baringthon_topological_2022} and \meths). The electronic band structure and energy band dispersion along $\overline{\text{M}}\overline{\Gamma}\overline{\text{M}}$
are plotted in the Suppl.~Mat.~\ref{S_HRARPES} showing a good agreement with ARPES data. The $\sigma_y$ s-DOS at the Fermi surface, originating from $S_1$ and $S_2$, and projected on the top surface (first BL, $n=1$) is displayed in Fig.~\ref{figSARPES}\reb{d}. The color code represents the $\sigma_y$-DOS $\mathcal{N}_\text{DOS}^{\sigma_y}$ projected onto the first BL. Around $\overline{\Gamma}$, positive (negative) values are observed for positive (negative) $k_x$. The sign of this Rashba field is opposite around the $\overline{\text{M}}$ point (close to 0.8~\ang$^{-1}$). Two electron ($P_1$, $P_3$) and one hole ($P_2$) pockets emerge, in very good agreement again with (S)ARPES data. By introducing the Rashba surface potential imposed by the surface symmetry breaking, we are able to reproduce the spin-resolved map over the projected 2D-Brillouin onto the first BiSb BL (see \meths).

\section{Ultrafast spin-charge conversion in B\lowercase{i}$_{1-x}$S\lowercase{b}$_x$/C\lowercase{o} probed by TH\lowercase{z}-TDS emission spectroscopy.}

\begin{figure*}[!htp]
  \begin{center}
      \includegraphics[width=\textwidth]{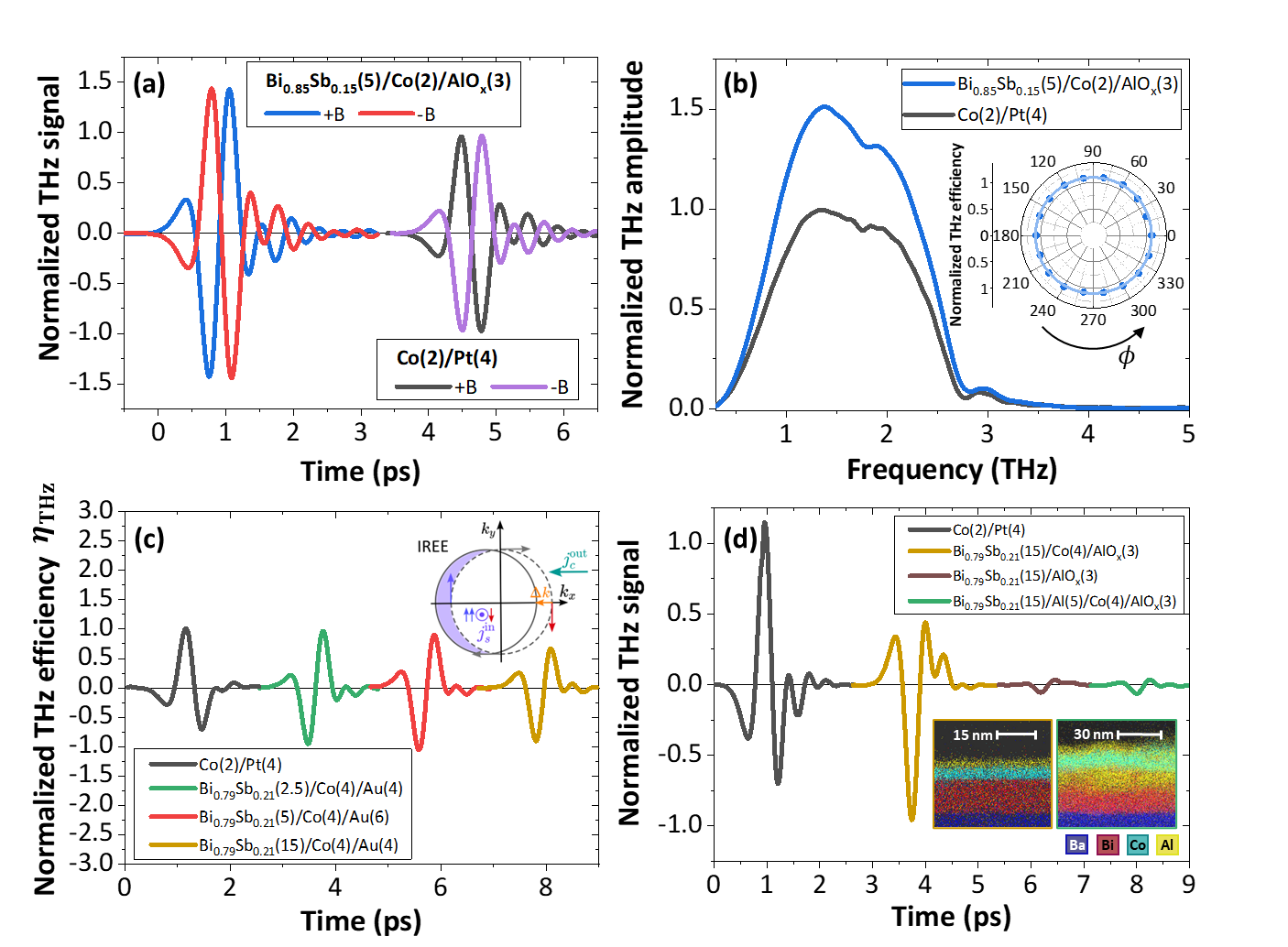}
       \caption{\textbf{SCC and THz emission from Bi$_{1-x}$Sb$_x$/Co bilayers.} (a) THz time-trace from Bi$_{0.85}$Sb$_{0.15}$(5)/Co(2)/AlO$_x$(3) for $\pm B$ compared to Co(2)/Pt(4) (grown on high resistivity Si substrates). THz phase reversal is a signature of SCC-mediated THz emission. (b) Spectral components of the THz emission from Bi$_{0.85}$Sb$_{0.15}$(5)/Co(2)/AlO$_x$(3) and Co(2)/Pt(4) for $+B$. Inset) Normalized THz efficiency $\eta_\text{THz}$ dependence on the azimuthal angle $\phi$ for Bi$_{0.85}$Sb$_{0.15}$(5)/Co(2)/AlO$_x$(3). (c) THz efficiency $\eta_\text{THz}$ as a function of the Bi$_{0.79}$Sb$_{0.21}$ layer thickness (2.5, 5 and 15~nm) compared to Co(2)/Pt(4) (high resistivity Si substrates). (d) THz signals from Bi$_{0.79}$Sb$_{0.21}$(15)/Al(5)/Co(4)/AlO$_x$(3), Bi$_{0.79}$Sb$_{0.21}$(15)/AlO$_x$(3), Bi$_{0.79}$Sb$_{0.21}$(15)/Co(4)/AlO$_x$(3) on BaF$_2$ (Suppl.~Mat.~\ref{S_substrate}) and Co(2)/Pt(4). Time traces are shifted in time for clarity. Inset) Fluorescence map obtained from a TEM cross-section for Bi$_{0.79}$Sb$_{0.21}$ (15)/Co(4)/AlOx(3) (brown frame) and Bi$_{0.79}$Sb$_{0.21}$(15)/Al(5)/Co(4)/AlOx(3) (green frame) grown on BaF$_2$ for the elements and the color code given below the maps.}
    \label{fig4}
  \end{center}
\end{figure*}


With the clear demonstration of the spin resolved surface states, we now discuss the dynamical spin-charge conversion in Bi$_{1-x}$Sb$_x$ capped by a thin Co layer as a spin-injector. THz emission spectroscopy in the time domain (THz-TDS) has recently emerged as a powerful spectroscopic technique to investigate ultrafast SCC in materials with strong SOC~\cite{Seifert2016} and, in particular, in TI/FM structures~\cite{Tong2021,Wang2018,Chen2021,Rongione2022}. The thin FM layer is excited by a femtosecond laser pump leading to ultrafast demagnetization. This generates both spin density $\hat{\mu}_s$ and spin-current $\mathcal{J}_s$ diffusing toward the BiSb/Co interface owing to the two spin populations and mobilities introduced by the \textit{sp}-band spin-splitting in Co~\cite{Dang2020}. The spin is afterwards converted into a transverse charge flow $\mathcal{J}_c$ on sub-picosecond timescales leading to a THz transient emission that is directly probed in the time-domain (see Refs.~\cite{Seifert2016,Dang2020} and \meths). Two SCC mechanisms can contribute: the IREE from the surface states and the ISHE from the bulk. The THz electric field can be expressed as:

\begin{equation}
\left\{
\begin{aligned}
\mathbf{E}^\text{ISHE}_\text{THz} (\theta) &\propto \frac{\partial \mathcal{J}_c^\text{ISHE}}{\partial t} \propto i\omega \theta_\text{SHE}~\left(\mathcal{J}_s \times \frac{\mathbf{M}(\theta)}{|\mathbf{M}|}\right)\\
\mathbf{E}^\text{IREE}_\text{THz} (\theta) &\propto \frac{\partial \mathcal{J}_c^\text{IREE}}{\partial t} \propto i\omega \Lambda^\text{IREE}~\left(\hat{\mu}_s \times \frac{\mathbf{M}(\theta)}{|\mathbf{M}|}\right) 
\end{aligned}
\right.
\label{E_scc}
\end{equation}
where $\mathbf{M}(\theta)$ is the Co layer magnetization vector,
$\hat{\mu}_s$ is the spin-accumulation vector relaxing on $S_1$ and $S_2$ and $\omega$ is the frequency. $\mathbf{M}(\theta)$ can be controlled by an external in-plane magnetic field $B$ at an angle $\theta$ from the $y$ axis. In the above equation, $\theta_\text{SHE}=\mathcal{J}_c/\mathcal{J}_s$ is the spin Hall angle scaling the bulk ISHE-mediated SCC, whereas $\Lambda^\text{IREE}$ is the Rashba-Edelstein length scaling the IREE from the surface states (refer to \meths~and Suppl.~Mat.~\ref{S_lin_resp_th}).

On Fig.~\ref{fig4}\reb{a}, we report the THz signal acquired in reflection geometry from Bi$_{0.85}$Sb$_{0.15}$(5)/Co(2) sample (numbers in parenthesis are thicknesses in nm) at room temperature with a saturating in-plane magnetic field $B\simeq \pm 100$~mT insuring that the magnetization follows the external field within much less than a degree. It is compared to the emission from our optimized Co(2)/Pt(4) metallic ISHE-type sample. In both cases, we observe a short picosecond THz pulse with some minor oscillations within a 3~ps wide envelope. The THz signal phase changes sign when reversing the magnetic field, in full agreement with SCC-mediated THz emission (Suppl.~Mat.~\ref{S_dipolar}). We note several differences: first, the amplitude from the BiSb/Co layer is 1.5 times larger than the Co/Pt revealing the large SCC efficiency in BiSb. For the same magnetic field orientation, the BiSb/Co THz signal phase is opposite to the one found in the Co/Pt sample (Fig.~\ref{fig4}\reb{a}): this sign inversion is related to the inverted layer stacking of the FM and non magnetic layer, giving an opposite phase and indicating that Pt and BiSb share the same conversion sign.

We also report in Fig.~\ref{fig4}\reb{b} the THz spectra (Fourier transform) obtained from Co(2)/Pt(4) and Bi$_{1-x}$Sb$_x$(5)/Co(2), demonstrating a relative power enhancement by a factor $\simeq2.3$ for the Bi$_{1-x}$Sb$_x$/Co sample (field enhancement by a factor $\times$1.5). The THz amplitude of Bi$_{1-x}$Sb$_x$/Co is also shown to scale linearly with the pump fluence (Suppl.~Mat.~\ref{S_fluence}), measured up to a few tens of \micro J.cm$^{-2}$. The azimuthal angular dependence of the THz emission obtained by rotating the sample in the plane by an angle $\phi$, while keeping fixed the magnetic field, is shown in the inset of Fig.~\ref{fig4}\textcolor{blue}{b}. The emission is almost isotropic revealing a pure SCC phenomenon with no evidence of non-linear optical effects such as shift or surge current contributions (\eg~photon drag, photogalvanic effects,\etc). This is in contrast with the recent report given by Park\etal\cite{park_topological_2022} where additional but small non-magnetic contributions were observed. At normal incidence of the optical pump pulse, a pure isotropic THz response \vs $\phi$ is expected from the linear response theory for both ISHE and IREE scenario (Suppl.~Mat.~\ref{S_lin_resp_th}) as discussed here. The role of the capping layer (metallic Au or oxidized Al, AlO$_x$) has been carefully excluded by control experiments (Suppl.~Mat.~\ref{S_capping}). Two additional control samples were grown: \textit{i}) a Bi$_{0.79}$Sb$_{0.21}$(15\,nm) sample free of ferromagnetic Co and only capped with naturally oxidized AlO$_x$(3\,nm) almost emitting no THz radiation (Fig.~\ref{fig4}\reb{d}) and \textit{ii}) a sample with a 5 nm thick Al metallic spacer inserted between BiSb and Co. The Al insertion strongly reduces the THz emission. It may be explained either by a larger near-infrared (NIR) and THz absorption in Al, or by interfacial spin-loss or weakening of the surface SCC induced by the strong degradation of the interface quality. Indeed, TEM pictures displayed in the inset of Fig.~\ref{fig4}\reb{d} reveal a strong intermixing of the Al interlayer that may induce a loss of an efficient spin-injection and/or the alteration of the surface states (more details in Suppl.~Mat.~\ref{S_interface}).


To get a better insight into the SCC mechanism, the thickness dependence of the THz efficiency, $\eta_\text{THz}$, is displayed in Fig.~\ref{fig4}\reb{c} for the Bi$_{0.79}$Sb$_{0.21}$ series (2.5, 5 and 15~nm thick layers).
$\eta_\text{THz}$ represents the spin-injection and conversion efficiencies and is obtained by withdrawing the NIR and THz absorptions in the heterostructure following the procedure proposed in Refs.~\cite{gueckstock2021,meinert2020,hawecker2022} (Suppl.~Mat.~\ref{S_efficiency}). Strikingly, $\eta_\text{THz}$ for BiSb remains constant for the whole thickness series where the BiSb bandgap widens dramatically by several hundreds of meV as the thickness is reduced, as calculated previously~\cite{baringthon_topological_2022}. Moreover, in agreement with our ARPES results~\cite{baringthon_topological_2022}, this demonstrates the absence of coupling between the top and bottom surface states~\cite{zhang_crossover_2010} down to 2.5~nm, unlike previously argued in the case of sputtered BiSb materials~\cite{sharma2021}. The characteristic evanescence length is indeed ultrashort, typically 2 BL (0.8~nm) near $\overline{\Gamma}$ as confirmed by density functional theory~\cite{ishida2016} for pure Bi and confirmed by our TB calculation 
(Suppl.~Mat.~\ref{S_lin_resp_th}). Such behavior is therefore more in favor of an interfacial origin of the SCC and thus strongly hints towards IREE from the surface states rather than ISHE from bulk.

\begin{figure*}[!htp]
  \begin{center}
      \includegraphics[width=\textwidth]{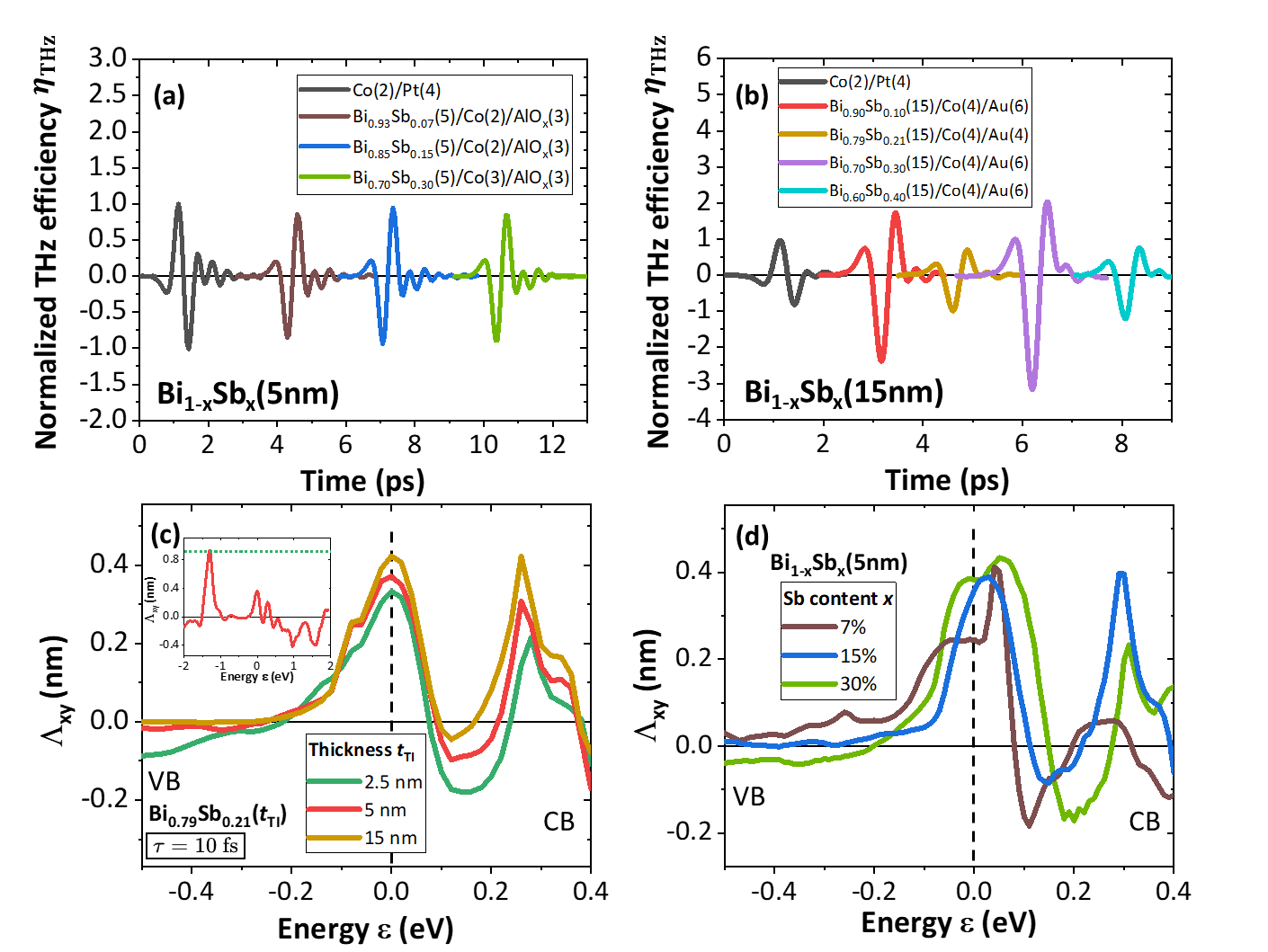}
       \caption{\textbf{Sb content dependence of the THz efficiency $\eta_\text{THz}$ and calculated IREE response for $\tau_s=10$~fs.} (a) $\eta_\text{THz}$ from Bi$_{1-x}$Sb$_x$(5)/Co(2-3)/AlO$_x$(3) with $x$=0.07, 0.15 and 0.3. (b) $\eta_\text{THz}$ from Bi$_{1-x}$Sb$_x$(15)/Co(4)/Au(4-6) with $x$=0.1, 0.21, 0.3 and 0.4. (c-d) Values of the IREE length $\Lambda_{xy}^{\text{IREE}}$ illustrating the conversion efficiency \vs energy $\varepsilon$ for (c) Bi$_{0.79}$Sb$_{0.21}$ layers with thicknesses 2.5, 5 and 15~nm and (d) Bi$_{1-x}$Sb$_x$(5nm) for $x=$~0.07, 0.15 and 0.3. The inset presents the full IREE response outside the bandgap in the bulk valence and conduction bands for Bi$_{0.79}$Sb$_{0.21}$(5nm). The calculations have been performed on for a spin-relaxation time $\tau_s=\tau=10$~fs after integration of the bands from to the top BL to the middle of the layer.}
    \label{fig5}
  \end{center}
\end{figure*}

Furthermore, when considering the possible contribution of \textit{bulk} ISHE, BiSb thickness has to be compared with its spin diffusion length. It has been evaluated in bulk BiSb at $\lambda_{sf}\simeq$~8~nm by Sharma\etal\cite{sharma2021}, which is much larger than 2.5~nm. If bulk states were to contribute \textit{via} ISHE, the spin current would therefore flow across the whole layer depth, and upper and lower interfaces would contribute similarly to the SCC but with an opposite sign. The net charge current would drop to zero, as $\tanh\left(\frac{t_{TI}}{\lambda_{sf}}\right)\tanh\left(\frac{t_{TI}}{2\lambda_{sf}}\right) \propto \frac{t_{TI}^2}{2\lambda_{sf}^2}$~\cite{Sinova2015} for small TI thicknesses $t_{TI}$ when considering the multiple spin current reflections at the TI interfaces~\cite{hawecker2022}. This is in contrast with the thickness-independent SCC observed here, in the ultrathin limit.
We thus anticipate that the bulk states are hardly involved, and that the surface states at the FM/TI interface are mainly responsible for a net charge current through IREE. This conclusion is also supported by \textit{i}) the large increase of the surface state DOS of BiSb at the interface compared to the DOS bulk states as the BiSb thickness increases (see Suppl.~Mat.~\ref{S_lin_resp_th}) and by \textit{ii}) the assumption of the preservation of BiSb surface states in exchange contact with Co as it seems to be the case.

We now focus on the THz efficiency $\eta_\text{THz}$ \vs the Sb content $x$, plotted for Bi$_{1-x}$Sb$_x$(5)/Co(2)/AlO$_x$(3)
in Fig.~\ref{fig5}\reb{a} and for Bi$_{1-x}$Sb$_x$(15)/Co(4)/Au
in Fig.~\ref{fig5}\reb{b} (Suppl.~Mat.~\ref{S_concentration}). In each series, a large $\eta_\text{THz}$ is measured, comparable to that of Co/Pt with a maximum for $x=0.3$ for the 15~nm series, \textit{i.e.} at the limit of the topological phase diagram~\cite{lenoir_chapter_2001}. Importantly, $\eta_\text{THz}$ remains similar for all the samples for the $t_\text{TI}=5$~nm series (Fig.~\ref{fig5}\reb{a}), highlighting again the prominent role of the surface states in the SCC. It also emphasizes the robustness of the IREE over a wide range of Sb content. The persistence of a large signal outside the bulk topological window ($x=0.4$ for 15~nm thickness, in Fig.~\ref{fig5}\reb{b}) indicates that bulk states may start to contribute by shunting a part of the spin current.

\section{SCC and IREE tensor from Linear Response theory}

We now compare the enhanced THz emission observed on ultrathin BiSb films to that obtained from the linear response theory described by the IREE response, namely $\Lambda_{xy}^{\text{IREE}}$ matching with the so-called inverse Edelstein length (see Suppl.~Mat.~\ref{S_lin_resp_th} and \meths). We consider an extended formalism to the one recently developed to address the direct REE response as given in Refs.~\cite{Fert2019,Johansson2021,Roy2022}. We evaluate here the \textit{intraband response} to an ongoing spin-current relaxing onto the Fermi surface generating, \textit{via} an out-of-equilibrium spin-density, a charge current according to: $\mathcal{J}_c^{x}=\Lambda_{xy}^{\text{IREE}}~ \mathcal{J}_{s,z}^{y}$ with $\Lambda_{xy}^{\text{IREE}}=\frac{\sum_{n,k}\left(\sigma^{y}_{nk}v_{nk}^x\tau_s\frac{\partial f_{nk}}{\partial \varepsilon}\right)}{\sum_{n,k}\left(\frac{\partial f_{nk}}{\partial \varepsilon}\right)}$
where $y$ is the direction of the spin injected with a flow along $z$ whereas $x$ is the direction of the in-plane charge current. In the above expression, $n$ is the band index, $v_{nk}^x$ the corresponding band velocity along $x$ and $\tau_s$ is the typical (spin) relaxation time assumed to be constant onto the Fermi surface. $f_{n\mathbf{k}}$ is the occupation number for the band $n$ and wavevector $\mathbf{k}$ whereas
$-\frac{\partial f_{n\mathbf{k}}}{\partial \varepsilon}=\mathcal{N}_\text{DOS}(\varepsilon,\mathbf{k},n)$ represents the local DOS in the $\textbf{k}$ space.

Calculations are performed on bare BiSb bilayers (BLs) free of any Co overlayer as investigated by SARPES. We show, on Fig.~\ref{fig5}\reb{c}, the energy dependence of $\Lambda_{xy}^{\text{IREE}}$ obtained for Bi$_{0.79}$Sb$_{0.21}$ of different film thicknesses: 2.5, 5 and 15~nm (see \meths). The Fermi level position corresponds to $\varepsilon=0$. 
One observes that $\Lambda_{xy}^{\text{IREE}}$ is largely enhanced in the bandgap region in the (-0.2 - 0.2)~eV window where the $S_2$ surface state and $S_1$ TSS are located, however without being able to differentiate their individual contributions (see Fig.~\ref{figS_IREEthick} for the corresponding DOS in the Suppl.~Mat.~\ref{S_lin_resp_th}). $\Lambda_{xy}^{\text{IREE}}$ remains constant \vs BiSb film thickness in this window range, as experimentally observed in THz data of Fig.~\ref{fig4}\reb{c}. Last, $\Lambda_{xy}^{\text{IREE}}$ is maximum at the Fermi level, where it reaches values in the range of the equivalent SCC efficiency of Pt, product of the spin-Hall angle by the spin-diffusion length $\lambda_{sf}$ as $\theta_\text{SHE}\times \lambda_{sf}\approx 0.2-0.3$~nm for the same spin or momentum relaxation time ($10$~fs). Increasing the spin-relaxation onto the surface states to 30~fs would yield $\Lambda_{xy}^{\text{IREE}}\approx 1$~nm.
$\Lambda_{xy}^{\text{IREE}}$ is displayed in Fig.~\ref{fig5}\reb{c} as a function of the Sb content $x$ for the 5~nm series. Although one observes a slight increase of the SCC response from $x$=0.1 to 0.3 in the gap window (still in line with the presence of the surface states, and possibly indicating a volume contribution), one may conclude that the SCC remains roughly constant at the vicinity of the Fermi energy ($\varepsilon=\varepsilon_\text{F}=0$). This is observed experimentally from THz-TDS measurements, and thus suggests an interfacial IREE nature of the SCC, at least for the thinner films. One cannot totally rule out a certain ISHE contribution arising from the bulk propagating bands associated to a very short spin-diffusion length. Nonetheless, the strong impedance mismatch and subsequent spin-backflow at Co/BiSb interface would be strongly in disfavor of such scenario.

\section*{Conclusions}


Although less investigated compared to other Bi-based families because of its modest bandgap,
BiSb ultrathin films still exhibit very robust surface states, as revealed by our spin-resolved ARPES measurements. This is in part related to the confinement effects without being detrimental to the surface states.  In particular, the presence of a topological surface state $S_1$ in a wide Sb-composition and thickness ranges is clearly observed, displaying helical spin texture at the Fermi surface with a specific opposite chirality between the electron pocket near the $\bar{\Gamma}$ point and the six hole pockets away from $\bar{\Gamma}$. \textit{Via} ultrafast THz emission spectroscopy, we demonstrate that the complex spin-texture gives rise to a very efficient spin-charge conversion, resulting from the spin-injection from a Co overlayer excited by an ultrashort laser pulse, mainly occurring \textit{via} inverse Rashba-Edelstein effect (IREE) owing to the strong localization of the surface states. Our results address the role of spin-textured hybridized Rashba-like surface states offering unprecedented SCC efficiency despite the breaking of the TRS symmetry due to the local exchange interactions imposed by the magnetic contact. These results hold promise for efficient and integrated structures based on BiSb. Future investigations will concern the dynamics of the spin relaxation onto the BiSb surface states excited by ultrashort pulses.

\section*{Acknowledgments}

 We thank T.~Kampfrath and G.~Bierhance (Freie Universität Berlin) for very fruitful discussions. This work was supported by a grant overseen by the French National Research Agency (ANR) as part of the “Generic Project Call - 2021” Programme (ANR-21-CE24-0011 TRAPIST), public grant overseen by the French National Research Agency (ANR) as part of the “Investissements d’Avenir” program (Labex NanoSaclay, reference: ANR-10-LABX-0035 SPICY) and the program ESR/EquipEx+ (Grant No.~ANR-21-ESRE-0025). We acknowledge financial support from the Horizon 2020 Framework Programme of the European Commission under FET-Open Grant No.~863155 (s-Nebula) and Grant No.~64735 (Extreme-IR).

\section*{Author contributions}

J.-M.G., H.J., P.LF., S.D. and A.L. conceived and designed the experiment. J.-M.G. supervised the project. L.B., D.S., Ma.Mo., A.L., N.R., F.B., P.LF. and J.-M.G. grew the samples and performed ARPES and SARPES experiments at Synchrotron Soleil. G.P, L.B. and A.L. performed TEM cross-section and SR-TEM microscopy. E.R, S.D., Ma.Mi., Ju.M., J.T. performed THz-TDS experiments at room temperature. H.J. performed TB calculations. E.R., P.LF., S.D., A.L., R.L., N. R., H.J, J.-M.G. analysed the data. E.R., P.LF., S.D., A.L., N. R., R.L., H.J., J.-M.G. wrote the paper.

\clearpage

\section*{Methods}
\label{meth}

\textbf{MBE growth of Bi$_{1-x}$Sb$_x$.} The Bi$_{1-x}$Sb$_x$ samples were grown by Molecular Beam Epitaxy (MBE). A Si(111) substrate was annealed in ultra-high vacuum (UHV) at 1370 K in order to obtain a 7$\times$7 surface reconstruction, observed by reflection high energy electron diffraction (RHEED). The Bi$_{1-x}$Sb$_x$ alloy is grown by co-deposition from two Knudsen cells. Each cell flux is calibrated prior to the deposition using a quartz microbalance and the relative deposition rates are directly used to estimate the Sb-concentration $x$. RHEED measurements are performed throughout the deposition. Up to 5 nm, we observe a 2D-growth; for thicker films, we stop the evaporation at 7 nm for a 10 min intermediate annealing at 500 K before completing the film growth up to the targeted thickness. This procedure is further described in Ref.~\cite{baringthon_topological_2022}. BiSb films with thicknesses from 2.5 to 15 nm and with various compositions (0.03<$x$<0.3) were grown using this method and their crystallographic quality was demonstrated by RHEED, X-ray-diffraction or Scanning Transmission Electron Microscopy (STEM)~\cite{baringthon_topological_2022}. \\
All our samples were grown in the MBE chamber of the CASSIOPEE beamline installed on the SOLEIL synchrotron. After their elaboration, they can be transferred in UHV to an ARPES (Angle-Resolved PhotoEmission Spectroscopy) or a SARPES (Spin-Resolved ARPES) experiments where we can characterize their electronic structure. After photoemission measurements, some of the sample were transferred again into the MBE chamber for Co electron beam deposition. These bilayer systems were used for SCC measurements using THz emission spectroscopy.

\vspace{0.1in}

\textbf{SEM - FIB milling and EDX.} Lamellae for STEM observation were prepared from the sample using Focused Ion Beam (FIB) ion milling and thinning. Prior to FIB ion milling, the sample surface was coated with 50 nm of carbon to protect the surface from the platinum mask deposited used for the ion milling process. Ion milling and thinning were carried out in a FEI SCIOS dual-beam FIB-SEM. Initial etching was performed at 30 keV, and final polishing was performed at 5 keV. The lamellae were prepared following the two different zone axis ($\langle 1 1 0 \rangle$ and $\langle 1 1 2 \rangle$) of the BaF$_2$ substrate. All samples were observed in an aberration-corrected FEI TITAN 200 TEM-STEM operating at 200 keV. The convergence half-angle of the probe was 17.6 mrad and the detection inner and outer half-angles for HAADF-STEM were 69 mrad and 200 mrad, respectively. All micrographs where 2048 by 2048 pixels. The dwell time was 8 \micro s and the total acquisition time 41 s. EDX measurements were performed in the Titan microscope featuring the Chemistem system, that uses a Bruker windowless Super-X four-quadrant detector and has a collection angle of 0.8 sr.

\vspace{0.1in}

\textbf{(Spin-)Angular-resolved photoemission spectroscopy.} The photoemission experiments were performed on the CASSIOPEE beamline installed on the SOLEIL storage ring (Saint-Aubin, France). The beamline hosts two endstations. A high-resolution ARPES endstation, which was used in this work for the measurement of the Fermi surface and of the band dispersion, using 20 eV incident photons with a linear horizontal polarization.  It is equipped with a Scienta R4000 electron analyzer. The photon spot size on the sample is of the order of 50$\times$50 \micro m$^2$ and the overall kinetic energy resolution (taking into account both the photon energy and the electron kinetic energy resolutions) was of the order of 10 meV.  The second endstation is a spin-resolved ARPES experiment, where the beam size is around 300$\times$300 \micro m$^2$. It is equipped with a MBS A1-analyzer with a 2D detector for ARPES measurements. Close to this 2D detector, a 1$\times$1 mm$^2$ hole collects photoelectron with well-defined kinetic energy and momentum. They are sent into a spin manipulator able to orient any spin component along the magnetization axis of a FERRUM VLEED spin-detector, made of a Fe(100)-p(1$\times$1)O surface~\cite{Bertacco99,ferrum} deposited on a W-substrate. 

The spin polarization along the selected direction is proportional to the difference of the two signals collected for opposite magnetizations of the Fe-oxide target. To reduce as much as possible the measurement asymmetries stemming from the instrument (\textit{i.e.}, not due to the spin polarization), four measurements per polarization direction are acquired, reversing both the ferrum magnetization direction, and the electron spin direction. This four measurements are combined into a geometrical average. The polarization is then determined by $P=S^{-1}(I_+^\sigma-I_-^\sigma)/(I_+^\sigma+I_-^\sigma)$.


The 1$\times$1 mm$^2$ hole introduces an integration on both the kinetic energy and the wave vector. For the kinetic energy, it corresponds to 0.23\% of the used pass energy (10 eV in our case), so around 23 meV. Convoluted with the energy resolution of the analyzer (10~meV for this pass energy and an entrance slit of 400~\micro m), it gives an overall kinetic energy resolution of 25 meV. For the wave vector, the 1~mm aperture corresponds to an integration on 4\% of the total (30$^\circ$) angular range, which gives 1.2$^\circ$. At 20 eV photon energy, for electrons at the Fermi level, this gives a $k$-resolution of around 0.048~\AA$^{-1}$. This explains the relatively broad features of the SARPES in Fig.~\ref{figSARPES}\reb{c,e} compared to Fig.~\ref{figSARPES}\reb{a}. \\
The analyzer optics is movable and can collect electrons in a large 2D (30$^\circ \times 30^\circ$) angular range. To map the spin texture at the Fermi level, the analyzer is set to the appropriate kinetic energy while the optics is moved by 0.2$^\circ$ step along two X and Y perpendicular directions. The two in-plane spin components are measured at each step.

\vspace{0.1in}

\textbf{THz emission spectroscopy.} Ultrafast  near-infrared (NIR) pulses ($\simeq$100 fs) centered at $\lambda_{\text{NIR}}$=810~nm are derived from a Ti:Sapphire oscillator to photo-excite the spin carriers directly from the front surface (Co side). Average powers of up to 600~mW were used with a repetition rate of 80~MHz (the energy per pulse is around 3~nJ). The typical laser spot size on the sample was about 200~\micro m~$\times$~200 \micro m. The optical pump is initially linearly polarized and irradiate the TI/FM heterostructure under normal incidence. The generated THz pulses were also collected from the front surface of the samples (\ie, reflection geometry) using a set of parabolic mirrors of 150 and 75~mm focal length to focus on the detection crystal. The samples were placed on a mount with a small magnetic field (around 100~mT) in the plane of the thin films. Both the sample orientation (angle $\phi$) and the in-plane magnetic field (angle $\theta$) can be independently rotated in the sample plane. Standard electro-optic sampling was used to detect the electric field of the THz pulses, using a 500~\micro m-thick $\langle 1\,1\,0 \rangle$ ZnTe crystal. A chopper was placed at the focal point between the second and third parabolic mirror to modulate the THz beam at 6 kHz for heterodyne lock-in detection. A mechanical delay line was used to sample the THz ultrafast pulse as a function of time. The THz propagation path was enclosed in a dry-atmosphere purged chamber (typically <2\% humidity) to reduce water absorption of the THz radiation.

\vspace{0.1in}

\textbf{Tight-binding calculations of Bi$_{1-x}$Sb$_x$ multilayers.} We have developed a tight binding (TB) model in order to describe the Bi$_{1-x}$Sb$_x$ electronic band structure as well as their surface topological properties~\cite{saito2016}. This approach is indeed well suited for TI and gives a fair description of the surface state spin texture in close agreement with the one derived from Density Functional Theory (DFT) developed for pure Bi surfaces~\cite{koroteev2004,hirahara2006,koroteev2008}. The rhombohedral A7 structure is described by two atoms per unit cell, forming then a bilayer (BL) of thickness of about 0.4~nm. Bi$_{1-x}$Sb$_x$ slabs are obtained by stacking the BL along the (1\,1\,1) direction ($z$ axis) with two different plane-to-plane distances. We constructed our Hamiltonian on the basis of the work of Ref.~\cite{saito2016} using the generalization of the \textit{sp$^3$} TB-model Hamiltonian proposed for bulk Bi and Sb crystals~\cite{liu1995}, adapted to Bi$_{1-x}$Sb$_x$ alloys~\cite{teo2008} and complemented by the introduction of additional surface potential terms when dealing with thin layers (treatment in slabs)~\cite{petersen2000,ast2012,saito2016}. In particular, the hopping parameters for the BiSb alloys are obtained by using the virtual crystal approximation (VCA) according to~\cite{teo2008}:

\begin{equation}
V_\text{C}^\text{BiSb}=x~V_\text{C}^\text{Sb}+(1-x^2)~V_\text{C}^\text{Bi}
\end{equation}
where $x$ is the antimony content and $V_\text{C}^\text{Sb}$ and $V_\text{C}^\text{Bi}$ are the respective hopping parameters of Sb and Bi taken from Ref.~\cite{liu1995}. One notes $\hat{\sigma}_\alpha$ the spin index on each atom where $\alpha$ stands for the directional index. The hopping terms among the atomic orbitals are decomposed into inter- and intra-BL hopping terms. The inter-BL off-diagonal hopping term between atoms (plane) 1 and atoms (plane) 2 consists of the nearest-neighbor coupling in the bulk BiSb Hamiltonian, whereas the intra-BL hopping term consists of two parts which represents respectively the third and second nearest neighbor contributions. We considered the overall TB Hamiltonian according to:

\begin{equation}
\hat{\mathcal{H}}=\hat{\mathcal{H}}_0+\hat{\mathcal{H}}_\text{SO}+\hat{\mathcal{H}}_\gamma
\end{equation}
where $\hat{\mathcal{H}}_0=\Sigma_{i\mu}^{j\nu} \ket{i\mu} V_{i\mu}^{j\nu}\bra{j \nu}$ represents the hopping Hamiltonian ($i$, $j$ are the atomic positions, $\mu$, $\nu$ are the orbitals), $\hat{\mathcal{H}}_\text{SO}=\frac{\hbar}{4m^2c^2}\left(\overrightarrow{\nabla} V(r) \times \hat{p}\right) \cdot \hat{\sigma}$ the SOC term and $\hat{\mathcal{H}}_\gamma$ the Rashba surface potential induced by the deformation of the surface orbitals due to the local electric field. Indeed, due to the symmetry breaking at the surface, a Rashba SOC term must be taken into account in the Hamiltonian at the two surface planes. We model such effect for the \textit{$sp^3$} basis by using the approach of Ast and Gierz~\cite{ast2012} for $\hat{\mathcal{H}}_\gamma$ considering two additional surface hopping terms $\gamma_{sp}$ and $\gamma_{pp}$ acting respectively between the $s-p_z$ and $p_x-p_z$ (or $p_y-p_z$) surface orbitals. We thus add the $\hat{\mathcal{H}}_\gamma$ Hamiltonian term of the form:

\begin{eqnarray}
\hat{\mathcal{H}_\gamma}=\left\{
\begin{array}{cc}
    \gamma_{sp} &  (i,i)\equiv (s,p_z) \\
    \pm \gamma_{sp1} &  (i,j)\equiv (s,p_z) \\
    \pm \gamma_{pp} \cos (\theta) & ~(i,j)\equiv (p_x,p_z) \\
    \pm \gamma_{pp} \sin (\theta) & ~(i,j)\equiv (p_y,p_z) \\
\end{array}
\right.
\end{eqnarray}
where the + (-) sign corresponds to the uppermost (lowermost) atomic plane and $\theta$ is the angle between the direction joining the two atoms considered and the $x$-direction. We then restrained ourselves to the in-plane surface hopping as for a pure 2D system. The best agreement with ARPES results is found by adding, as proposed in Ref.~\cite{ast2012}, additional on-site $s-p_z$ coupling $\gamma_{sp}=-0.2$~eV, and surface hopping terms $\gamma_{sp1}$=0.3~eV and $\gamma_{pp}$=-0.6~eV for $x$=0.15 slightly departing from the values given for pure Bi, \ie~$\gamma_{sp}$=0.45~eV and $\gamma_{pp}$=-0.27~eV~\cite{saito2016}, with opposite sign for the top and bottom surfaces due to the opposite direction of the potential gradient. We emphasize that this surface terms are required to correctly reproduce the surface state dispersion as observed by ARPES experiments. The size of the Hamiltonian $\hat{\mathcal{H}}(k_x,k_y)$ to diagonalize is $16N \times 16N$ where $N$ is the number of bilayers (BLs). Once the Green function of the multilayer system is defined as:

\begin{equation}
    \hat{G}(\varepsilon,k_x,k_y)=\left[\varepsilon+i\delta -\hat{\mathcal{H}}(k_x,k_y)\right]^{-1}
\end{equation}

the partial density of state (DOS) $\mathcal{N}_\text{DOS}(\varepsilon)$ \vs the energy $\varepsilon$ equals $\mathcal{N}(n,\varepsilon)=-(1/\pi)~\text{Im} \Tr[\hat{G}(\varepsilon,n,k_x,k_y)]$ whereas the spin density of states (spin-DOS) with spin along the $\alpha$ direction is $s_\alpha(\varepsilon)=(1/\pi) ~\text{Im} \Tr[\hat{\sigma}_\alpha \hat{G} (\varepsilon,k_x,k_y)]$. $\delta$ is the typical energy broadening ($\simeq$ 10 meV) and the trace ($\Tr$) is applied over the considered $sp^3$ orbitals on a given BL index ($n\in [1,N]$). The energy zero ($\varepsilon$=0) refers to the Fermi level position.

\vspace{0.1in}

The modelling of IREE were performed by TB method. We have summed the contributions from the different Fermi surface pockets within the 2D-BZ after \textit{i}) having introduced the Rashba potentials at the BiSb surfaces, required to match both the TSS electronic dispersion and SML measured in (S)ARPES experiments~\cite{baringthon_topological_2022}; and after \textit{ii}) having considered a same (spin) relaxation time of $\tau_s=\tau_0=10$~fs involved in the intraband transitions. 
The inverse Edelstein length has been then evaluated on the whole Fermi surface according to the following expression $\Lambda_{xy} (\varepsilon) =  \frac{\sum_{{n}} \int d^2\mathbf{k} \bra{\psi_{n\mathbf{k}}} \hat{v}_x\tau_0 \ket{\psi_{n\mathbf{k}}} \bra{\psi_{n\mathbf{k}}} \hat{\sigma}_y\ket{\psi_{n\mathbf{k}}}\mathcal{N}_\text{DOS}(\varepsilon,\mathbf{k},n) \tau}{\sum_{{n}} \int d^2\mathbf{k} \mathcal{N}_\text{DOS}(\varepsilon,\mathbf{k},n)}$ (see Suppl. Mat.~\ref{S_lin_resp_th}).

The typical value of $\tau_s=10$~fs corresponds to an energy broadening $\Gamma=\hbar/(2\tau_s) \simeq 50$~meV) and a Fermi velocity of about $5\times 10^{5}$~m.s$^{-1}$ like extracted from ARPES data. The trace on $\kappa_{xy}$ is performed by summing the BL contributions from the top surface down to the middle of the BL, typically matching the typical finite TSS evanescence or extension length. 

\end{sloppypar}

\bibliographystyle{ieeetr}
\bibliography{biblio}

\end{document}


\title{Supplementary Material of "Spin-momentum locking and ultrafast spin-charge conversion in ultrathin epitaxial Bi$_{1-x}$Sb$_x$ topological insulator."}
\date{\today}
\author{E.~Rongione}
\affiliation{Unité Mixte de Physique, CNRS, Thales, Université Paris-Saclay, F-91767 Palaiseau, France}
\affiliation{Laboratoire de Physique de l’Ecole Normale Supérieure, ENS, Université PSL, CNRS, Sorbonne Université,
Université Paris Cité, F-75005 Paris, France}
\author{L.~Baringthon}
\affiliation{Unité Mixte de Physique, CNRS, Thales, Université Paris-Saclay, F-91767 Palaiseau, France}
\affiliation{Synchrotron SOLEIL, L’Orme des Merisiers, Départementale 128, F-91190 Saint-Aubin, France}
\affiliation{Université Paris-Saclay, CNRS, Centre de Nanosciences et de Nanotechnologies, F-91120 Palaiseau, France}
\author{D.~She}
\affiliation{Unité Mixte de Physique, CNRS, Thales, Université Paris-Saclay, F-91767 Palaiseau, France}
\affiliation{Synchrotron SOLEIL, L’Orme des Merisiers, Départementale 128, F-91190 Saint-Aubin, France}
\affiliation{Université Paris-Saclay, CNRS, Centre de Nanosciences et de Nanotechnologies, F-91120 Palaiseau, France}
\author{G.~Patriarche}
\affiliation{Université Paris-Saclay, CNRS, Centre de Nanosciences et de Nanotechnologies, F-91120 Palaiseau, France}
\author{R.~Lebrun}
\affiliation{Unité Mixte de Physique, CNRS, Thales, Université Paris-Saclay, F-91767 Palaiseau, France}
\author{A.~Lema\^{\i}tre}
\affiliation{Université Paris-Saclay, CNRS, Centre de Nanosciences et de Nanotechnologies, F-91120 Palaiseau, France}
\author{M. Morassi}
\affiliation{Université Paris-Saclay, CNRS, Centre de Nanosciences et de Nanotechnologies, F-91120 Palaiseau, France}
\author{N.~Reyren}
\affiliation{Unité Mixte de Physique, CNRS, Thales, Université Paris-Saclay, F-91767 Palaiseau, France}
\author{M.~Mičica}
\affiliation{Laboratoire de Physique de l’Ecole Normale Supérieure, ENS, Université PSL, CNRS, Sorbonne Université,
Université Paris Cité, F-75005 Paris, France}
\author{J.~Mangeney}
\affiliation{Laboratoire de Physique de l’Ecole Normale Supérieure, ENS, Université PSL, CNRS, Sorbonne Université,
Université Paris Cité, F-75005 Paris, France}
\author{J.~Tignon}
\affiliation{Laboratoire de Physique de l’Ecole Normale Supérieure, ENS, Université PSL, CNRS, Sorbonne Université,
Université Paris Cité, F-75005 Paris, France}
\author{F. Bertran}
\affiliation{Synchrotron SOLEIL, L’Orme des Merisiers, Départementale 128, F-91190 Saint-Aubin, France}
\author{S.~Dhillon}
\affiliation{Laboratoire de Physique de l’Ecole Normale Supérieure, ENS, Université PSL, CNRS, Sorbonne Université,
Université Paris Cité, F-75005 Paris, France}
\affiliation{\normalfont Corresponding authors:~\href{mailto:sukhdeep.dhillon@phys.ens.fr}{sukhdeep.dhillon@phys.ens.fr},~\href{mailto:henri.jaffres@cnrs-thales.fr}{henri.jaffres@cnrs-thales.fr},~\href{mailto:jeanmarie.george@cnrs-thales.fr}{jeanmarie.george@cnrs-thales.fr}}
\author{P.~Le Fèvre}
\affiliation{Synchrotron SOLEIL, L’Orme des Merisiers, Départementale 128, F-91190 Saint-Aubin, France}
\author{H.~Jaffrès}
\affiliation{Unité Mixte de Physique, CNRS, Thales, Université Paris-Saclay, F-91767 Palaiseau, France}
\affiliation{\normalfont Corresponding authors:~\href{mailto:sukhdeep.dhillon@phys.ens.fr}{sukhdeep.dhillon@phys.ens.fr},~\href{mailto:henri.jaffres@cnrs-thales.fr}{henri.jaffres@cnrs-thales.fr},~\href{mailto:jeanmarie.george@cnrs-thales.fr}{jeanmarie.george@cnrs-thales.fr}}
\author{J.-M.~George}
\affiliation{Unité Mixte de Physique, CNRS, Thales, Université Paris-Saclay, F-91767 Palaiseau, France}
\affiliation{\normalfont Corresponding authors:~\href{mailto:sukhdeep.dhillon@phys.ens.fr}{sukhdeep.dhillon@phys.ens.fr},~\href{mailto:henri.jaffres@cnrs-thales.fr}{henri.jaffres@cnrs-thales.fr},~\href{mailto:jeanmarie.george@cnrs-thales.fr}{jeanmarie.george@cnrs-thales.fr}}

\maketitle

\begin{sloppypar}
\small
This Supplementary Material presents the structural properties of Bi$_{1-x}$Sb$_x$ derived from TEM and EDX characterization (\ref{S_struct}), the electronic properties of Bi$_{1-x}$Sb$_x$ derived from ARPES (\ref{S_HRARPES}), the THz emission from Bi$_{1-x}$Sb$_x$/Co as a function of various parameters: the applied magnetic field (\ref{S_dipolar}), the fluence dependence of the THz emission (\ref{S_fluence}), the THz transients renormalization procedure (\ref{S_efficiency}), the Bi$_{1-x}$Sb$_x$ thickness dependence of the renormalized THz emission (\ref{S_thickness}), the substrate influence on the THz emission (\ref{S_substrate}), the THz renormalized emission as a function of the Bi$_{1-x}$Sb$_x$ alloy concentration (\ref{S_concentration}), as well as control experiments: the THz spin-injection efficiency depending on the interface quality (\ref{S_interface}) or the influence of the non-magnetic capping (\ref{S_capping}), and finally presents the discussions about the spin-charge conversion profile via interfacial TSS contributions and linear response theory (\ref{S_lin_resp_th}).

\setcounter{figure}{0}
\setcounter{table}{0}
\setcounter{section}{0}

\newpage

\section{Structural and electronic properties.}
\label{S_struct}

\begin{figure}[!htp]
  \begin{center}
      \includegraphics[width=\textwidth]{figS_struct.png}
       \caption{\textbf{Structural properties of epitaxial Bi$_{1-x}$Sb$_x$ thin films on BaF$_2$ substrate.} (a) Quantitative EDX-STEM linescan profile analysis of the heterostructures (BaF$_2$/Bi$_{0.79}$Sb$_{0.21}$(15nm)/Co(4nm)/AlO$_x$(3nm)/Carbon protective layer). (b) HAADF-STEM image of Bi$_{0.79}$Sb$_{0.21}$(15nm)/Co(4nm)/AlO$_x$(3nm) sample. The BaF$_2$ substrate is oriented following a $\langle 1 1 0 \rangle$ zone axis. The green dotted box represents the area studied in EDX-STEM chemical analysis. (c) EDX-STEM elemental mapping of Bi$_{0.79}$Sb$_{0.21}$(15nm)/Co(4nm)/AlO$_x$(3nm) (on BaF$_2$ substrate) composed with Bi, Co, Al and Ba elements. (d) Graphical representation of the Bi$_{1-x}$Sb$_x$ bilayer decomposition, with the presence of intra-bilayer and inter-bilayer distances.}
    \label{fig1}
  \end{center}
\end{figure}

We present in Fig.~\ref{fig1} the structural description of epitaxially grown BiSb ultrathin films. The sample studied is Bi$_{0.79}$Sb$_{0.21}$(15nm)/Co(2nm)/AlO$_x$(3nm) grown either onto BaF$_2$ and SiHR substrates. We report on Fig.~\ref{fig1}\reb{a-b} the EDX-STEM sectional picture of the sample grown onto BaF$_2$ displaying the Bi, Ba, Co and Al elements. We can observe that the interfaces between \textit{i}) the substrate and Bi$_{1-x}$Sb$_x$ as well as \textit{ii}) Bi$_{1-x}$Sb$_x$ and Co are very smooth, demonstrating also the good epitaxial quality of the deposited BiSb layers. We performed complementary and quantitative EDX-STEM profile analysis along a linescan corresponding to the layer thickness. We map a very weak interdiffusion of the chemical species. These are reported on Fig.~\ref{fig1}\reb{c}. For comparison, the Bi$_{1-x}$Sb$_x$ unit cell is graphically represented in Fig.~\ref{fig1}\reb{d}, where each Bi$_{1-x}$Sb$_x$ bilayer (1 BL $\simeq$ 0.4 nm) is defined with an inter-distance about 0.3 nm and a intra-bilayer distance about 0.1 nm.\\

\begin{figure}[!htp]
  \begin{center}
      \includegraphics[width=0.8\textwidth]{figS_structSi.png}
       \caption{\textbf{Structural properties of epitaxial Bi$_{1-x}$Sb$_x$ thin films on Si substrate.} (a) HAADF-STEM image of Bi$_{0.79}$Sb$_{0.21}$(15nm)/Co(4nm)/AlO$_x$(3nm) sample on Si substrate. (b) EDX-STEM elemental mapping of Bi$_{0.79}$Sb$_{0.21}$(15nm)/Co(4nm)/AlO$_x$(3nm) (on Si substrate) composed with Bi, Co, Al and Si elements.}
    \label{figSi_TEM}
  \end{center}
\end{figure}

Alongside BaF$_2$ substrate, we have also developed the growth of Bi$_{1-x}$Sb$_x$ topological insulator on Si substrate which structural properties are shown in Fig.~\ref{figSi_TEM}. Fig.~\ref{figSi_TEM} shows the EDX-STEM images obtains when growing Bi$_{0.79}$Sb$_{0.21}$(15nm)/Co(4nm)/AlO$_x$(3nm) on Si(111) substrate and attesting of the good quality of the MBE. We observe that the interfaces composing the heterostructures, whether it is the Si/Bi$_{1-x}$Sb$_x$ interface or the Bi$_{1-x}$Sb$_x$/Co interface, are well defined and very smooth. This demonstrates the good epitaxial quality of the thin films realized on Si substrates. In particular, the quality of the Co/BiSb interface is very similar to that when the growth is on BaF$_2$ substrates. More details can be found in the PhD thesis of L.~Baringthon~\cite{baringthon_elaboration_2022} and in Ref.~\cite{baringthon_topological_2022}.

\section{High resolution spin-resolved ARPES: spin texture of B\lowercase{i}$_{x}$S\lowercase{b}$_{1-x}$ surface states.}
\label{S_HRARPES}


\begin{figure}[!htp]
  \begin{center}
      \includegraphics[width=.8\textwidth]{FS1.pdf}
       \caption{\textbf{ARPES and TB results obtained from 5~nm Bi$_{1-x}$Sb$_x$ ($x$=0.15).}  (a) ARPES measurements along the $\overline{\Gamma}\overline{\text{M}}$ direction showing the energy band dispersions of two surface states $S_1$ and $S_2$ and of the \textit{bulk} valence bands. (b) DOS TB-based calculations of the corresponding structure along the $\overline{\Gamma}\overline{\text{M}}$ line of the Brillouin zone and projected onto the first BL. (c) Corresponding spin-resolved DOS calculated by the TB method and projected onto the first BL. (d) Plot of both DOS and spin-resolved DOS projected onto the first BL showing the large spin-polarization along $\overline{\Gamma}\overline{\text{M}}$.}
    \label{figHRARPES}
  \end{center}
\end{figure}

Fig.~\ref{figHRARPES}\reb{a} displays the ARPES measurements acquired on a 5~nm (12 BLs) Bi$_{0.85}$Sb$_{0.15}$ at room temperature. The figure displays respectively the DOS-resolved projected onto the Fermi surface (top) and the energy dispersions obtained along the $\overline{\Gamma}\overline{\text{M}}$. This emphasizes, in particular, the energy band dispersions of two surface states $S_1$ (colored in red) and $S_2$ (colored in blue) as well as the \textit{bulk} valence bands (4 bands are clearly visible).

Fig.~\ref{figHRARPES}\reb{b} displays the resulting 2D-DOS for the same 12 BLs Bi$_{0.85}$Sb$_{0.15}$ obtained by TB methods showing a very good agreement with ARPES data. Fig.~\ref{figHRARPES}\reb{c} displays the corresponding spin-resolved DOS showing large values in the same range than the DOS previously calculated. The energy splitting appears clearly in particular near the $\overline{\Gamma}$ point (DOS and spin-resolved DOS are given in arbitray units). 

Fig.~\ref{figHRARPES}\reb{d} shows the respective values of the 2D-DOS (blue) and 2D spin-resolved DOS (red) at the Fermi level along the $\overline{\Gamma}-\overline{\text{M}}$ line. The different peaks corresponds to the occurrence of the $S_1$ and $S_2$ resonances at the Fermi level (band crossed) and responsible for the SCC. One notes that the spin-DOS are asymmetric \vs the $\overline{\Gamma}$ point as expected emphasizing on the particular chirality of those Rashba-like states. One notes that the values of the spin-DOS represents a large fraction of the total DOS (between 50\% and 100\%) which corresponds to average values for $\bra{\psi_{\mathbf{k}n}}\hat{\sigma}_y\ket{\psi_{\mathbf{k}n}}$ or the order of $\bra{\psi_{\mathbf{k}n}}\hat{\sigma}_y\ket{\psi_{\mathbf{k}n}}\approx \pm (0.5-0.7)~ \frac{\hbar}{2}$.


\clearpage

\section{TH\lowercase{z} emission from B\lowercase{i}$_{1-x}$S\lowercase{b}$_x$.}

\subsection{TH\lowercase{z} emission \textit{vs.} the magnetic field and dipolar emission.}
\label{S_dipolar}

We present on Fig.~\ref{figS_mag} the THz emission obtained from Bi$_{0.85}$Sb$_{0.15}$(5)/Co(2) as a function of the magnetic field angle $\mathbf{B}(\theta)$ in polar plot. The blue region indicates a positive polarity (or phase) of the THz timetrace whereas the red color indicates a negative phase in agreement with the convention of Fig.~\ref{fig4} of the main text. We recover then the expected dipolar emission (with a characteristic sinus shape) owing to the cross product $\mathcal{J}_c\propto (\mathcal{J}_s,\mu_s)\times \mathbf{M}$ between the spin current $j_s$ or spin accumulation (spin density) $\mu_s$ and the in-plane magnetization $\mathbf{M}(\theta)$ (convoluted with the detection direction of the electro-optic detector). The presence of symmetrical emission lobes indicates an absence of additional contributions to the THz emission (\ie non-magnetic contributions).

\begin{figure}[!htp]
  \begin{center}
      \includegraphics[width=0.6\textwidth]{figS_mag.png}
       \caption{\textbf{THz emission from Bi$_{1-x}$Sb$_x$(5, $x$=0.15)/Co(2)/AlO$_x$(3) as a function of the magnetic field angle $\mathbf{B}(\theta)$.} The magnetic field is applied in the sample plane with an angle $\theta$ from the $y$-axis. The measurements have been performed at room temperature (RT).}
    \label{figS_mag}
  \end{center}
\end{figure}

\subsection{Fluence dependence on the TH\lowercase{z} emission.}
\label{S_fluence}

We report on Fig.~\ref{figS_fluence} the fluence dependence of Bi$_{0.93}$Sb$_{0.07}$(5nm)/Co(2nm)/AlO$_x$(3nm) for laser pump power ranging from 0 to 350 mW. We observe a linear dependence of the type $E_\text{THz}\propto P_\text{NIR}$ in agreement with the absorption/excitation process and subsequent spin-charge conversion ~\cite{Seifert2016}. No saturation of the THz amplitude with the pump fluence could be observe, we thus place our study below both damage and saturation thresholds.

\begin{figure}[!htp]
  \begin{center}
      \includegraphics[width=0.6\textwidth]{figS_fluence.png}
       \caption{\textbf{Fluence dependence on the THz emission.} THz peak-peak amplitude from Bi$_{0.93}$Sb$_{0.07}$(5nm)/Co(2nm)/AlO$_x$(3nm) on SiHR as a function of the pump power and fluence. The measurements have been performed at RT.}
    \label{figS_fluence}
  \end{center}
\end{figure}

\subsection{TH\lowercase{z} renormalization procedure by withdrawing the optical absorptions.}
\label{S_efficiency}

We followed the method proposed in Refs.~\cite{gueckstock2021,meinert2020,hawecker2022}. We extract, from the measured THz electric field $\mathbf{E}_\text{THz}$, the THz efficiency $\eta_\text{THz}$ directly linked to the spin-charge conversion and spin-injection properties according to:

\begin{equation}
\abs{\mathbf{E}_{\text{THz}}} \propto \left( \frac{Z_0 \times \mathcal{A}_\text{NIR}}{1+n_{\text{sub}}(\omega)+Z_0 \sum _i \sigma_{i}(\omega)t_\text{i}} \right) \times \left(\alpha_\text{SCC} ~\eta_{\text{THz}}\right)
\label{eff_prod}
\end{equation}
where $Z_0=377$~\Ohm~is the free-space impedance, $\sigma_i=1/\rho_i$ is the layer conductivity (obtained from the resistivity $\rho_i$), $t_i$ is the layer thickness, $\alpha_\text{SCC}$ is the spin-charge conversion efficiency and $n_\text{sub}$ is the THz refractive index of the substrate (presently $n_\text{sub}^{\text{BaF}_2}=2.7$~\cite{querry_optical_1987} and $n_\text{sub}^{\text{Si}}=3.2$, considered constant over the studied frequency range from 0.3 to 3 THz). The near-infrared (NIR) radiation absorption in thin films is modelled via Beer-Lambert absorption $\mathcal{A}_\text{NIR} = \prod_i \exp(-\alpha_i t_i)$ where $\alpha_i$ is the absorption coefficient. The considered values for the studied heterostructures are collected in Table~\ref{tabthz_NIR}.

\begin{table}[!htp]
\begin{tabular}{|c|c|c|}
\cline{1-3}
    \textbf{Material} & \textbf{Resistivity $\rho$~(\micro\Ohm.cm)} & \textbf{Absorption coefficient $\alpha$ (cm$^{-1}$)} \\ \cline{1-3}
     \hline\hline
    Co & 29 & $7.5 \times 10^5$ \cite{palik1998} \\ \cline{1-3}
    Pt & 30 & $7.7 \times 10^5$ \cite{palik1998} \\ \cline{1-3}
    Au & 25 & $9.7 \times 10^5$ \cite{werner2009} \\ \cline{1-3}
    W & 25 & $4.3 \times 10^5$ \cite{palik1998} \\ \cline{1-3}
    Al & 40 & $1.3 \times 10^6$ \cite{palik1998}\\ \cline{1-3}
    AlO$_x$ & - & $1.15 \times 10^6$ \cite{Rakic95} \\ \cline{1-3}
    Bi$_{1-x}$Sb$_x$ & 560 & - \\ \cline{1-3}
\end{tabular}
\caption{Considered values for the resitivities $\rho$ and the absorption coefficients $\alpha$ (at pump wavelength $\lambda_\text{NIR}$=800 nm) depending on the material. Resistivity values are obtained from magneto-transport measurements.}
\label{tabthz_NIR}
\end{table}

\subsection{Renormalization of the TH\lowercase{z} emission as a function of the B\lowercase{i}$_{1-x}$S\lowercase{b}$_x$ thickness.}
\label{S_thickness}

In this section, we present the THz emission and renormalized THz efficiency  $\eta_\text{THz}$ for two thickness-dependent series, either \textit{i}) $t_\text{TI}=\{2.5,5,15\}$ nm for $x=0.21$ (presented in the main text in Fig.~\ref{fig4}\reb{c}) and \textit{ii}) $t_\text{TI}=\{5,50\}$ nm for $x=0.15$. In the first case whose results are displayed in Fig.~\ref{figS_renorm1}, we present the THz emission and THz efficiency $\eta_\text{THz}$ for Bi$_{1-x}$Sb$_x$($t_\text{TI}$)/Co(4nm)/Au(4-6nm) grown on SiHR substrate. We observe an almost independent THz efficiency as a function of the Bi$_{1-x}$Sb$_x$ thickness, in favor of an interfacially-mediated spin-charge conversion process. The efficiency is similar to the Co(2)/Pt(4) reference.

\begin{figure}[!htp]
  \begin{center}
      \includegraphics[width=0.9\textwidth]{figS_renorm2.png}
       \caption{\textbf{Renormalization procedure of the THz emission for the thickness-dependent series.} (a) THz emission and (b) THz efficiency of Bi$_{1-x}$Sb$_x$($t_\text{TI}$)/Co(4nm)/Au(4-6nm) grown on SiHR substrate as a function of Bi$_{1-x}$Sb$_x$ thickness from $t_\text{TI}=\{2.5, 5, 15\}$ nm (and Sb content $x=0.21$). \textit{Time traces are shifted in time for clarity.}}
    \label{figS_renorm1}
  \end{center}
\end{figure}

Secondly, we present on Fig.~\ref{figS_renorm2} the THz signal from Bi$_{0.85}$Sb$_{0.15}$($t_\text{TI}$)/Co(2)/AlO$_x$(3nm) (second Sb content series grown on SiHR substrate) with $t_\text{TI}$ ranging from 5 to 50 nm. The THz emission from 5~nm-thick sample reach up to $\times 1.5$ the emission from reference Co(2)/
Pt(4). Both signal and THz efficiency $\eta_\text{THz}$ of the 50 nm-thick sample is reduced compared to the 5 nm-thick Bi$_{1-x}$Sb$_x$ sample, respectively by a factor 2.2 and 1.9.

\begin{figure}[!htp]
  \begin{center}
      \includegraphics[width=0.9\textwidth]{figS_thickness2.png}
       \caption{\textbf{Bi$_{1-x}$Sb$_x$ thickness dependence on the THz emission.} (a) Measured THz signals and (b) renormalized THz efficiency $\eta_\text{THz}$ from Bi$_{0.85}$Sb$_{0.15}$(5nm)/Co(2nm)/AlO$_x$(3nm) and Bi$_{0.85}$Sb$_{0.15}$(50nm)/Co(2nm)/AlO$_x$(3nm) compared to Co(2nm)/Pt(4nm). The substrate is SiHR. \textit{Time traces are shifted in time for clarity.}}
    \label{figS_renorm2}
  \end{center}
\end{figure}

\subsection{Substrate dependence on the TH\lowercase{z} emission from B\lowercase{i}$_{1-x}$S\lowercase{b}$_x$/C\lowercase{o} bilayers.}
\label{S_substrate}

Fig.~\ref{figS_substrate} displays the impact of the substrate on both THz emission and THz efficiency $\eta_\text{THz}$ for identical Bi$_{0.79}$Sb$_{0.21}$(15nm)/Co(4nm)/AlO$_x$(3nm) (or Au(4nm)) structures. The THz signal is larger on BaF$_2$ substrates; however a larger THz efficiency $\eta_\text{THz}$ is observed on SiHR substrates on which we mainly focus in the present work. This is explained by the difference in THz refractive index between the substrates~\cite{querry_optical_1987}. Thus, in the remaining part of the study, we exclude the use of BaF$_2$ substrate as it can lead to \textit{i}) non-linear optical contributions to the THz emission and to \textit{ii}) charging effects in (S)ARPES experiments.

\begin{figure}[!htp]
  \begin{center}
      \includegraphics[width=0.9\textwidth]{figS_substrate.png}
       \caption{\textbf{Impact of the substrate on the THz emission.} (a) THz emission and (b) THz efficiency $\eta_\text{THz}$ of Bi$_{0.79}$Sb$_{0.21}$(15nm)/Co(3nm)/AlO$_x$(3nm) (or Au(4nm)) as a function of the substrate (either BaF$_2$, Si or SiHR) compared to Co(2nm)/Pt(4nm). \textit{Time traces are shifted in time for clarity.}}
    \label{figS_substrate}
  \end{center}
\end{figure}

\subsection{Renormalization of the TH\lowercase{z} emission as a function of the alloy ratio B\lowercase{i}$_{1-x}$S\lowercase{b}$_x$.}
\label{S_concentration}

In this section, we present the raw THz emission (prior to any renormalization treatment) and efficiency $\eta_\text{THz}$ (after renormalization) from Bi$_{1-x}$Sb$_x$/Co alloy as a function of the Sb concentration for the two sample series: \textit{i}) Bi$_{1-x}$Sb$_{x}$(15)/Co(4) for Sb concentration $x=\{0.1, 0.21, 0.3, 0.4\}$ (in Fig.~\ref{figS_concentration1}) and \textit{ii}) Bi$_{1-x}$Sb$_{x}$(5)/Co(2) for Sb concentration $x=\{0.07,0.15,0.3\}$ (in Fig.~\ref{figS_concentration2}). Surprisingly, we observe that a sizeable THz contribution is recovered at $x=0.4$, typically out of the topological window~\cite{benia_surface_2015,lenoir_chapter_2001}, thus possibly addressing the role of spin-textured Rashba-like hybridized surface states to the overall THz spin-charge conversion.

\begin{figure}[!htp]
  \begin{center}
      \includegraphics[width=\textwidth]{figS_concentration.png}
       \caption{\textbf{Bi$_{1-x}$Sb$_x$ alloy dependence.} (a) THz emission and (b) THz efficiency of Bi$_{1-x}$Sb$_x$(15)/Co(4)/Au as a function of the Sb concentration, presently $x=\{0.1,0.21,0.3,0.4\}$ compared to Co(2)/Pt(4). \textit{Time traces are shifted in time for clarity.}}
    \label{figS_concentration1}
  \end{center}
\end{figure}

\begin{figure}[!htp]
  \begin{center}
      \includegraphics[width=\textwidth]{figS_concentration2.png}
       \caption{\textbf{Bi$_{1-x}$Sb$_x$ alloy dependence.} (a) Measured THz emission and (b) THz efficiency of Bi$_{1-x}$Sb$_x$(5)/Co(2)/AlO$_x$(3) as a function of the Sb concentration, presently $x=\{0.07,0.15,0.30\}$ compared to Co(2)/Pt(4). \textit{Time traces are shifted in time for clarity.}}
    \label{figS_concentration2}
  \end{center}
\end{figure}

\section{Additional TH\lowercase{z} experiments on controlled samples: Absence of self-emission, effect of interface degradation and capping.}
\label{S_capping}

\begin{figure}[!htp]
  \begin{center}
      \includegraphics[width=\textwidth]{figS_int.png}
       \caption{\textbf{Interfacially-mediated THz emission from Bi$_{1-x}$Sb$_x$/Co.} (a) THz emission collected on Bi$_{1-x}$Sb$_x$/Co bilayer, Bi$_{1-x}$Sb$_x$/AlO$_x$ and Bi$_{1-x}$Sb$_x$/Al/Co/Al compared to Co/Pt reference. (b-c) Transmission electron microscopy (TEM) images in case of (b) Bi$_{1-x}$Sb$_x$/Co/Al and (c) Bi$_{1-x}$Sb$_x$/Al/Co/Al samples. The interface is strongly degraded in the latter due to Al spacer intermixing, leading to possible interfacial SCC destruction. \textit{Time traces are shifted in time for clarity.}}
    \label{figS_interface}
  \end{center}
\end{figure}

\subsection{Interface degradation and TH\lowercase{z} emission.}
\label{S_interface}

In order to investigate the origin of the spin-charge conversion and THz generation, we performed complementary measurements on two additional samples: one sample free of any ferromagnetic Co (only capped with naturally oxidized AlO$_x$(3)) and one sample with a metallic Al(5) spacer adding between the TI Bi$_{1-x}$Sb$_x$(15) and the ferromagnetic Co(4). We shown in Fig.~\ref{figS_interface} their THz emission compared to the metallic Co/Pt reference and the emission from Bi$_{1-x}$Sb$_x$/Co. Clearly, owing to the THz signal level from the bare BiSb sample free of Co, the THz emission in Bi$_{1-x}$Sb$_x$/Co mainly originates from SCC at Co/BiSb interfaces. Moreover, once a thin metallic Al spacer is inserted between Bi$_{1-x}$Sb$_x$ and Co, the resulting emission is strongly reduced at the same level of the bare  Bi$_{1-x}$Sb$_x$ sample. Several hypothesis would explain this emission reduction: \textit{i}) NIR and THz absorptions in the Al spacer, \textit{ii}) degradation of the spin-injection efficiency and/or \textit{iii}) a reduction of the interfacial spin-charge conversion efficiency. In more details, from Table~\ref{tabthz_NIR}, we can estimate the differential absorption from an additional Al(5) layer that we evaluate at  $\mathcal{A}_\text{NIR}^{'} \simeq 0.52$ and $\mathcal{Z}_\text{THz}^{'} \simeq 0.7$ respectively for the NIR and THz radiations. It leads to a down-scaling of the THz emission from Bi$_{1-x}$Sb$_x$/Al/Co/Al of about $\times$3 while experimentally a reduction by a factor $\times$14 is observed. Then a major part of the signal difference (factor $\times$5) originates either from a loss of the spin-injection and/or a loss of spin-charge conversion efficiencies.

\subsection{Impact of the capping on the TH\lowercase{z} emission from B\lowercase{i}$_{1-x}$S\lowercase{b}$_x$/C\lowercase{o} bilayers.}
\label{S_capping}

To exclude the contributions from the capping layers (AlO$_x$, Au and Co), we performed reference measurements on Co(2)/AlO$_x$(3) and Co(4)/Au(4) as presented in Fig.~\ref{figS_capping}. We observe that both signals are small in amplitude compared to the Co/Pt reference, about an order of magnitude smaller. The small THz contribution could arise from the anomalous Hall effect in Co.

\begin{figure}[!htp]
  \begin{center}
      \includegraphics[width=0.55\textwidth]{figS_capping.png}
       \caption{\textbf{Capping reference THz measurements.} THz emission from Co(2)/AlO$_x$(3) and Co(4)/Au(4) normalized with respect to Co(2)/Pt(4). \textit{Time traces are shifted in time for clarity.}}
    \label{figS_capping}
  \end{center}
\end{figure}




\clearpage

\section{Linear response theory for Rashba-Edelstein effect.}
\label{S_lin_resp_th}

In this section, we develop the linear response theory to describe the Inverse Rashba-Edelstein response from the generalized Kubo formula applied to BiSb ultrathin films. We treat first \textit{i}) the tight-binding modelling of BiSb films before tackling \textit{ii}) the calculations of the inverse Rashba-Edelstein tensor obtained from the linear response theory, before evidencing \textit{iii}) the (crystallographic) angular response of the spin-charge conversion to address the IREE contribution via the surface states of Bi$_{1-x}$Sb$_x$, as discussed previously in Ref.~\cite{Rongione2022} and in agreement with (S)ARPES measurements.

\subsection{Linear response theory in electronic transport.}

We present here the linear response theory and principle of the Rashba-Edelstein response. Generally, one searches for the response function of a given operator $\hat{\mathcal{O}}$ under a certain excitation described by an interaction Hamiltonian $\hat{\mathcal{H}}_{int}=\int \hat{\mathcal{B}} F(t) d^3r dt$ where $\hat{\mathcal{B}}$ is a quantum operator and $F(t)$ is a classical source. One can then divide the response function $\mathcal{R}$ with $<\hat{\mathcal{O}}>(\omega)=\mathcal{R} F(\omega)$ ($\omega$ is the frequency) \textit{via} respective \textit{intraband} and \textit{interband} terms from the respective Fermi sea and Fermi surface contributions. One obtain for $\omega=0$ in the limit of low frequency (Drude model): 

\begin{eqnarray}
\mathcal{R}^{I(a)}=\frac{e^2}{2\pi V} \Tr<\hat{\mathcal{O}} G^R(\epsilon_\text{F})\hat{\mathcal{B}} G^A(\varepsilon_\text{F})>\\
\mathcal{R}^{I(b)}=-\frac{e^2}{4\pi V} \Tr<\hat{\mathcal{O}} G^R(\varepsilon_\text{F})\hat{\mathcal{B}} G^R(\varepsilon_\text{F})+ c.c. >\\
\mathcal{R}^{II}=\frac{e^2}{4\pi V} \int_{-\infty} ^{+\infty} f(\varepsilon) d\varepsilon 
\Tr<\left[\hat{\mathcal{O}}  G^R \hat{\mathcal{B}}\frac{\partial G^R}{\partial \varepsilon}-\hat{\mathcal{O}}  \frac{\partial G^R}{\partial \varepsilon} \hat{\mathcal{B}}G^R \right]+ c.c. >
\end{eqnarray}
\vspace{0.1in}
where $G^{R,A}=\frac{1}{\varepsilon-\hat{\mathcal{H}}\pm i\Gamma}$ are the respective retarded ($R$) and advanced ($A$) Green's functions. The intraband response is given by the sum of the two terms, $\mathcal{R}^{I(a)}+\mathcal{R}^{I(b)}$ with $\mathcal{R}^{I(a)}$ generally considered as much larger for clean systems whereas the interband term (Fermi sea) is given by $\mathcal{R}^{II}$.

\subsubsection{Intrinsic Inverse spin Hall effect from the linear response theory.~\cite{litvinov2020,marder2000}}

For ISHE, we are searching for the expression of the 2D in-plane current density $\mathcal{J}_c$ along the $x$ direction when a spin-current is injected. The interaction term is $\hat{\mathcal{H}}_{int}=\int \hat{\mathcal{J}_s} E(t) d^3r dt$ where $E$ is the electric field and $\hat{\mathcal{J}_s}$ is the spin current. Starting from the Kubo formula, we obtain:

\begin{eqnarray}
\sigma_{xy}^{I(a)}=\frac{e^2}{2\pi V} \Tr<\hat{v}_y G^R(\varepsilon_\text{F})\hat{v}_xG^A(\varepsilon_\text{F})>\\
\sigma_{xy}^{I(b)}=-\frac{e^2}{4\pi V} \Tr<\hat{v}_y G^R(\varepsilon_\text{F})\hat{v}_xG^R(\varepsilon_\text{F})+ c.c. >\\
\sigma_{xy}^{II}=\frac{e^2}{4\pi V} \int_{-\infty} ^{+\infty} f(\varepsilon) d\varepsilon \Tr<\left[\hat{v}_y G^R \hat{v}_x \frac{\partial G^R}{\partial \varepsilon}-\hat{v}_y \frac{\partial G^R}{\partial \varepsilon} \hat{v}_x G^R \right]+ c.c. >
\end{eqnarray}

We write the intrinsic contribution to the spin-Hall conductivity after having introduced the velocity operator $\hat{\mathbf{v}} = \partial \hat{\mathcal{H}}/ \partial \mathbf{p}$, and the spin current operator $\hat{\mathcal{J}}_{zy}=\{\hat{\mathbf{v}}_z \hat{\sigma}_y\}$ according to:

\begin{equation}
    \sigma_{xz}^{y} = \frac{e^2 \hbar}{2\pi} \sum_{{n~;~m \ne n}} \int_{\text{BZ}}  \left(f_n(\mathbf{k}) -f_m(\mathbf{k})\right)\frac{\left[\bra{\psi_{n\mathbf{k}}} \hat{v}_x \ket{\psi_{m\mathbf{k}}} \bra{\psi_{m\mathbf{k}}} \{\hat{\sigma}_y\hat{v}_z\} \ket{\psi_{n\mathbf{k}}} \right]}{(\varepsilon_{n\mathbf{k}} - \varepsilon_{m\mathbf{k}})^2+\Gamma^2} \frac{d^2 \mathbf{k}}{(2\pi)^2}
\end{equation}
where $e$ the electron charge, $\hbar$ the reduced Planck constant, $\Gamma=\hbar/2\tau$ the typical energy broadening, $\ket{n}$ and $\ket{m}$ the quantum states for the wavevector $\mathbf{k}$. $\psi_{i\mathbf{k}}$ is the electronic state associated with energy $\varepsilon_{i\mathbf{k}}$ and $f_i(\mathbf{k})$ is the Fermi filling factor. 

\vspace{0.1in}

Note however that the ISHE gives zero contribution to the charge current for non-propagating states along the direction normal to the layers that is the case in the bandgap of the BiSb layers.

\subsubsection{Inverse Rashba-Edelstein tensor from the linear response theory.}

We have adapted this formalism to the following interaction term $\hat{\mathcal{H}}_{int}=-\int \hat{\mathcal{M}} \dot{B}(t) d^2r dt$ where $\hat{\mathcal{M}}=\frac{e\hbar}{m^*}\hat{\sigma}$ and $B$ is the magnetic field according to the formalism developed by Shen\etal\cite{shen2014}. We have then expanded this framework to the inverse Rashba-Edelstein tensor $\Lambda_{xy}^{\text{IEE}}=\mathcal{J}_c / \mathcal{J}_s$ where $\mathcal{J}_c $ is the 2D in-plane charge current along $x$ whereas $\mathcal{J}_s=\mathcal{J}_{sz}^{y}$ is the 3D ongoing spin-current with flux directed along the direction $z$ normal to the layer carrying a spin $\sigma_y$. Starting from the Kubo formula and retaining the $I(a)$ term only (intraband or Fermi surface term) for the intrinsic spin-Hall conductivity described here for clean epitaxial samples: 

We find respectively the local charge-current $\mathcal{J}_{c\mathbf{k}}$ and local magnetization density $\mu^{y}_{\mathbf{k}}$ generated by $\dot{B}_y$:

\begin{align}
    \mathcal{J}_{ck}(\varepsilon)&=\frac{e^2\hbar^2}{2\pi m^*\Gamma}  \sum_{{n}} f_{\mathbf{k}n}(\varepsilon)\frac{\left[\bra{\psi_{n\mathbf{k}}} \hat{v}_x \ket{\psi_{n\mathbf{k}}} \Gamma \bra{\psi_{n\mathbf{k}}} \hat{\sigma}_y\ket{\psi_{n\mathbf{k}}} \right]}{(\varepsilon - \varepsilon_{n\mathbf{k}})^2+\Gamma^2} \dot{B}\nonumber\\
    &=\frac{e^2\hbar }{m^*} \sum_{{n}} f_{\mathbf{k}n}(\varepsilon) \bra{\psi_{n\mathbf{k}}} \hat{v}_x \tau_{\mathbf{k}n}\ket{\psi_{n\mathbf{k}}} \bra{\psi_{n\mathbf{k}}} \hat{\sigma}_y\ket{\psi_{n\mathbf{k}}}\mathcal{N}_\text{DOS}(k,\varepsilon) \dot{B}\\
    \mathcal{\mu}_{k}(\varepsilon)&=\frac{e\hbar^2}{2\pi m^*\Gamma}  \sum_{{n}} f_{\mathbf{k}n}(\varepsilon)\frac{\left[\bra{\psi_{n\mathbf{k}}} \hat{\sigma}_y \ket{\psi_{n\mathbf{k}}} \Gamma \bra{\psi_{n\mathbf{k}}} \hat{\sigma}_y\ket{\psi_{n\mathbf{k}}} \right]}{(\varepsilon - \varepsilon_{n\mathbf{k}})^2+\Gamma^2} \dot{B} \nonumber\\
    &=\frac{e\hbar}{m^*}  \sum_{{n}} f_{\mathbf{k}n}(\varepsilon)\bra{\psi_{n\mathbf{k}}} \hat{\sigma}_y \ket{\psi_{n\mathbf{k}}} \bra{\psi_{n\mathbf{k}}} \hat{\sigma}_y\ket{\psi_{n\mathbf{k}}}\mathcal{N}_\text{DOS}(k,\varepsilon) \tau_{\mathbf{k}n}\dot{B}
\end{align}
where $\tau_{\mathbf{k}n}=\tau_0 \bra{\psi_{n\mathbf{k}}} \hat{\sigma}_y \ket{\psi_{n\mathbf{k}}}^2$ is the characteristic spin-relaxation time at point $\mathbf{k}$ for spin oriented along the $y$ direction injected onto the subband $n$ in the Brillouin zone. This specific time $\tau_\mathbf{k}$ ($\mathbf{k}$-dependent) scales with characteristic time $\tau_0$ (the mobility time) and takes into account the possible fast-relaxation at point $\mathbf{k}$ when the spin-direction ($y$) does not match with the corresponding eigenvector spin direction of the band. Spin-current $\mathcal{J}_{s\mathbf{k}}^y$ and $\mu_{\mathbf{k}}^y$ are linked by $\mu_\mathbf{k}^y=\mathcal{J}_{s\mathbf{k}}^y \tau_\mathbf{k}$ whereas $\mathcal{J}_{s\mathbf{k}}^y=\frac{\sum_n \mathcal{N}_\text{DOS}(\varepsilon,\mathbf{k},n)}{\sum_n \int \mathcal{N}_\text{DOS}(\varepsilon,\mathbf{k},n) d^2 k} \mathcal{J}_{s}$. We thus obtain:

\begin{eqnarray}
   \mathcal{J}_{c}^x(\varepsilon)= \frac{\sum_{{n}} \int d^2\mathbf{k} \bra{\psi_{n\mathbf{k}}} \hat{v}_x\tau_0 \ket{\psi_{n\mathbf{k}}} \bra{\psi_{n\mathbf{k}}} \hat{\sigma}_y\ket{\psi_{n\mathbf{k}}}\mathcal{N}_\text{DOS}(\varepsilon,\mathbf{k},n) }{\sum_{{n}} \int d^2\mathbf{k} \mathcal{N}_\text{DOS}(\varepsilon,\mathbf{k},n)  } \mathcal{J}_{sz}^y
\end{eqnarray}
which defines thus the IREE tensor $\Lambda_{xy}$ at the same time that the average IREE length.

\begin{equation}
    \Lambda_{xy}^{\text{IEE}} (\varepsilon) =  \frac{\sum_{{n}} \int d^2\mathbf{k} \bra{\psi_{n\mathbf{k}}} \hat{v}_x\tau_0 \ket{\psi_{n\mathbf{k}}} \bra{\psi_{n\mathbf{k}}} \hat{\sigma}_y\ket{\psi_{n\mathbf{k}}}\mathcal{N}_\text{DOS}(\varepsilon,\mathbf{k},n)}{\sum_{{n}} \int d^2\mathbf{k} \mathcal{N}_\text{DOS}(\varepsilon,\mathbf{k},n) } 
\end{equation}

We note that for the case of topological states represented by a single dispersion cone corresponding to the band $n$, $\Lambda_{xy}$ reduces to $\Lambda_{xy}=\bra{\psi_{n\mathbf{k}}} \hat{v}_x\tau_0 \ket{\psi_{n\mathbf{k}}} \bra{\psi_{n\mathbf{k}}} \hat{\sigma}_y\ket{\psi_{n\mathbf{k}}}$ as expected. 

Projecting the in-plane wavevector $\mathbf{k}:=(k \cos{(\phi-\phi_\text{R})},k \sin{(\phi-\phi_\text{R})})$ onto a particular crystallographic direction (\textit{e.g.} $\phi_\text{R}=0 \Leftrightarrow \overline{\Gamma \text{M}}$ in the reciprocal space), one finally obtains:

\begin{equation}
\begin{split}
    \Lambda_{xy}^{\text{IEE}} (\theta,\phi_\text{R}) &= \frac{\hbar^3}{ 2\pi m^*} \int_\text{BZ} d\phi~ \hat{v}_x (\phi,\phi_\text{R}) \hat{\sigma}_y (\phi,\theta,\phi_\text{R}) \sum_{n} \frac{ f_n(k) d^2 k}{\Gamma^2} 
    \\
    &= \frac{\hbar^3}{2\pi m^*\Gamma} \int_{0}^{2\pi} d\phi~ \hat{v}_x (\phi,\phi_\text{R}) \hat{\sigma}_y (\phi,\theta,\phi_\text{R}) \left( \frac{d^2 k}{d \varepsilon} \right) 
    \\
    &= \frac{\hbar^3}{2 \pi m^* \Gamma} \int_0^{2\pi} d\phi~ \hat{v}_x (\phi,\phi_\text{R}) \hat{\sigma}_y (\phi,\theta,\phi_\text{R}) \mathcal{N}_\text{DOS} (\phi,\phi_\text{R})
\end{split}
\end{equation}
where we have introduced the density of states $\mathcal{N}_\text{DOS} (\varepsilon)=(d^2 k / d \varepsilon)$, $\theta$ the orientation of out-of-equilibrium injected spins (in-plane magnetization or magnetic field), $\phi_\text{R}$ the in-plane sample orientation and $\phi$ the generic azimuthal angle. Therefore, the conversion symmetries are given by the integrated product between the longitudinal group velocity, the spin orientation, and the density of states. For $C_{3v}$ symmetry systems, the energy dispersion relation including Rashba term \cite{Fu2009,Vajna2012,Johansson2016} reads:

\begin{equation}
\varepsilon_\pm (\mathbf{k}) = \frac{\hbar^2 k^2}{2 m} \pm \sqrt{\alpha_\text{R}^2 k^2 + \lambda^2 k^6 \cos^2 (3 \phi)}
\end{equation}
with $\alpha_\text{R}$ the Rashba coefficient (velocity) and $\lambda$ a coefficient describing the strength of the six-fold anisotropy term. Thus we assume the spin accumulation, group velocity and density of states to have the following symmetries:

\begin{eqnarray}
\begin{split}
v_x (\phi,\phi_\text{R}) &\equiv \cos(\phi) + 3 \lambda \cos(3(\phi-\phi_\text{R})) \left[ \cos(\phi_\text{R}) \cos(\phi-\phi_\text{R}) + \sin(\phi_\text{R}) \sin(\phi-\phi_\text{R}) \right]
\\
\sigma_y(\phi,\theta,\phi_\text{R}) &\equiv \frac{1}{2} \left( 1 + \sin(\phi + \phi_\text{M}) \right) \times (1 + 3 \beta \cos^2(3 (\phi - \phi_\text{R})))
\\
\mathcal{N}_\text{DOS} (\phi,\phi_\text{R}) &\equiv 1 + \gamma \cos^2 (3 (\phi - \phi_\text{R}))
\end{split}
\end{eqnarray}
where $\lambda, \beta, \gamma$ are three coefficients describing the anisotropy (hexagonal wrapping) of each involved physical quantity. One obtains the general isotropic response $\Lambda_{xy}$:

\begin{equation}
\Lambda_{xy} (\theta, \phi_\text{R}) = \Lambda_{xy} (\theta)=\frac{\pi}{32} (3 \beta(8+18 \lambda+3 \gamma (2+5 \lambda))+2(8+12 \lambda +\gamma (4+9 \lambda))) \sin(\theta)
\end{equation}
independent of the crystallographic orientation $\phi_\text{R}$. This confirms our results of isotropic THz emission. This fulfills the dipolar symmetry for the THz emission \textit{vs.}~the in-plane magnetic field direction.

\subsubsection{Comparing IREE and ISHE.}

IREE is scaled by the characteristic IREE length $\Lambda_\text{IEE}$ whereas the ISHE is scaled by the spin-Hall conductivity (SHC) $\sigma_\text{SHC}=\sigma_{xz}^{y}$. In order to compare the efficiency of the two different processes, one must compare $\Lambda_\text{IEE}$ to the product $\theta_\text{SHE}\lambda_{sf}$ where $\theta_\text{SHE}=\sigma_\text{SHC}\times \rho_{xx}$ refers to the effective spin-Hall angle and $l_{sf}$ is the spin-diffusion length. One can thus transform $\Lambda_\text{IEE}$ to an equivalent SHC according to $\Lambda_\text{IEE}=\sigma_\text{IEE}\rho_{xx}\times \lambda_{sf}=\sigma_\text{IEE} r_s$ with $r_s=\frac{\tau_s}{e^2\mathcal{N}_\text{DOS}^\text{3D}\lambda_{sf}}$ the spin resistance.

We arrive at the following formula $\sigma_\text{IEE}=\frac{1}{2\sqrt{3}\pi^2}\left(\frac{e^2}{\hbar}\right) k_\text{F}^2 \Lambda_\text{IEE}$ to compare with $\sigma_\text{SHE}$ if non-zero.

\subsection{TB electronic band structure of Bi$_{1-x}$Sb$_x$.}

\subsubsection{Method.}

We have developed a tight binding (TB) model in order to describe the Bi$_{1-x}$Sb$_x$ electronic band structure as well as their surface topological properties~\cite{saito2016}. This approach is indeed well suited for TI and gives a fair description of the surface state spin texture in close agreement with the one derived from Density Functional Theory (DFT) developed for pure Bi surfaces~\cite{koroteev2004,hirahara2006,koroteev2008}. The rhombohedral A7 structure is described by two atoms per unit cell, forming then a bilayer (BL) of thickness of about 0.4~nm. Bi$_{1-x}$Sb$_x$ slabs are obtained by stacking the BL along the (1\,1\,1) direction ($z$ axis) with two different plane-to-plane distances. We constructed our Hamiltonian on the basis of the work of Ref.~\cite{saito2016} using the generalization of the \textit{sp$^3$} TB-model Hamiltonian proposed for bulk Bi and Sb crystals~\cite{liu1995}, adapted to Bi$_{1-x}$Sb$_x$ alloys~\cite{teo2008} and complemented by the introduction of additional surface potential terms when dealing with thin layers (treatment in slabs)~\cite{petersen2000,ast2012,saito2016}. In particular, the hopping parameters for the BiSb alloys are obtained by using the virtual crystal approximation (VCA) according to~\cite{teo2008}:

\begin{equation}
V_\text{C}^\text{BiSb}=x~V_\text{C}^\text{Sb}+(1-x^2)~V_\text{C}^\text{Bi}
\end{equation}
where $x$ is the antimony content and $V_\text{C}^\text{Sb}$ and $V_\text{C}^\text{Bi}$ are the respective hopping parameters of Sb and Bi taken from Ref.~\cite{liu1995}. One notes $\hat{\sigma}_\alpha$ the spin index on each atom where $\alpha$ stands for the directional index. The hopping terms among the atomic orbitals are decomposed into inter- and intra-BL hopping terms. The inter-BL off-diagonal hopping term between atoms (plane) 1 and atoms (plane) 2 consists of the nearest-neighbor coupling in the bulk BiSb Hamiltonian, whereas the intra-BL hopping term consists of two parts which represents respectively the third and second nearest neighbor contributions. We considered the overall TB Hamiltonian according to:

\begin{equation}
\hat{\mathcal{H}}=\hat{\mathcal{H}}_0+\hat{\mathcal{H}}_\text{SO}+\hat{\mathcal{H}}_\gamma
\end{equation}
where $\hat{\mathcal{H}}_0=\Sigma_{i\mu}^{j\nu} \ket{i\mu} V_{i\mu}^{j\nu}\bra{j \nu}$ represents the hopping Hamiltonian ($i$, $j$ are the atomic positions, $\mu$, $\nu$ are the orbitals), $\hat{\mathcal{H}}_\text{SO}=\frac{\hbar}{4m^2c^2}\left(\overrightarrow{\nabla} V(r) \times \hat{p}\right) \cdot \hat{\sigma}$ the spin-orbit term and $\hat{\mathcal{H}}_\gamma$ the equivalent-Rashba surface potential induced by the deformation of the surface orbitals due to the local electric field. Indeed, due to the symmetry breaking at the surface, additional hopping interaction terms must be taken into account in the Hamiltonian at the two surface planes. We model such effect for the \textit{$sp^3$} basis by using the approach of Ast and Gierz~\cite{ast2012} for $\hat{\mathcal{H}}_\gamma$ considering three additional surface hopping terms: on-site $\gamma_{sp_z}$ hopping, and neighboring hopping terms  $\gamma_{sp}^1=\gamma_{sp}^2$ and $\gamma_{pp}$ acting respectively between the $s-p_z$ and $p_x-p_z$ (or $p_y-p_z$) surface orbitals. We thus add the $\hat{\mathcal{H}}_\gamma$ Hamiltonian term of the form:

\begin{eqnarray}
\hat{\mathcal{H}_\gamma}=\left\{
\begin{array}{cc}
    \pm \gamma_{pp} \cos (\theta) & ~(i,j)\equiv (p_x,p_z) \\
    \pm \gamma_{pp} \sin (\theta) & ~(i,j)\equiv (p_y,p_z) \\
    \pm \gamma_{sp} &  (i,j)\equiv (s,p_z) \\
\end{array}
\right.
\end{eqnarray}
where the + (-) sign corresponds to the uppermost (lowermost) atomic plane and $\theta$ is the angle between the direction joining the two atoms considered and the $x$-direction. We then restrained ourselves to the in-plane surface hopping as for a pure 2D system. The best agreement with ARPES results is found by adding, as proposed in Ref.~\cite{ast2012}, additional on-site $s-p_z$ coupling $\gamma_{sp}=-0.2$~eV, and surface hopping terms $\gamma_{sp1}$=0.3~eV and $\gamma_{pp}$=-0.6~eV for $x$=0.15 slightly departing from the values given for pure Bi, \ie~$\gamma_{sp1}$=0.45~eV and $\gamma_{pp}$=-0.27~eV~\cite{saito2016}, with opposite sign for the top and bottom surfaces due to the opposite direction of the potential gradient. We emphasize that this surface terms are required to correctly reproduce the surface state dispersion as observed by ARPES experiments. The size of the Hamiltonian $\hat{\mathcal{H}}(k_x,k_y)$ to diagonalize is $16N \times 16N$ where $N$ is the number of bilayers (BLs). Once the Green function of the multilayer system is defined as:

\begin{equation}
    \hat{G}(\varepsilon,k_x,k_y)=\left[\varepsilon+i\delta -\hat{\mathcal{H}}(k_x,k_y)\right]^{-1}
\end{equation}

The partial density of state (DOS) $\mathcal{N}_\text{DOS}(\varepsilon)$ \vs the energy $\epsilon$ equals $\mathcal{N}(n,\varepsilon)=-(1/\pi)~\text{Im} \Tr[\hat{G}(\varepsilon,n,k_x,k_y)]$ whereas the spin density of states (spin-DOS) with spin along the $\alpha$ direction is $s_\alpha(\varepsilon)=(1/\pi)~\text{Im} \Tr[\hat{\sigma}_\alpha \hat{G} (\varepsilon,k_x,k_y)]$. $\delta$ is the typical energy broadening ($\simeq$~10 meV) and the trace ($\Tr$) is applied over the considered $sp^3$ orbitals on a given BL index ($n\in [1,N]$). The energy zero ($\varepsilon$=0) refers to the Fermi level position.

\subsubsection{Bi$_{1-x}$Sb$_x$ energy dispersion: surface states vs. Bloch states.}

\begin{figure}[!htp]
  \begin{center}
      \includegraphics[width=\textwidth]{FigS12.pdf}
       \caption{\textbf{Bi$_{0.79}$Sb$_{0.21}$ energy dispersion as a function of the thickness.} Tight-binding calculations of the dispersion profile of Bi$_{1-x}$Sb$_x$ as a function of the thickness $t_\text{TI}$ ranging from (a) 2.5 nm (6 BLs), (b) 5 nm (12 BLs) and (c) 15 nm (36 BLs).}
    \label{figS_TB_thick_dispersion}
  \end{center}
\end{figure}

Fig.~\ref{figS_TB_thick_dispersion} displays the typical electron energy dispersions obtained for Bi$_{0.79}$Sb$_{0.21}$ (respectively 2.5 (a), 5 (b) and 15~nm (c)) and corresponding spin-resolved DOS along the $\overline{\Gamma \text{M}}$ direction projected onto the first top BL. This emphasizes on the typical spin-splitting arising on the two $S_1$ and $S_2$ surface states. The spin-resolved DOS plotted in colorcode (red for positive value and blue for negative values) represents a large fraction of the total DOS (typically between 50\% and 70\% corresponding to average typical values $\bra{\psi_{\mathbf{k}n}}\sigma_y\ket{\psi_{\mathbf{k}n}}\simeq \pm 0.5-0.7 \hbar$). The bulk valence and electronic bands are also shown, displaying a specific spin-resolved texture but however much less pronounced.

\begin{figure}[!htp]
  \begin{center}
      \includegraphics[width=\textwidth]{figS_evanescence.png}
       \caption{\textbf{Evanescence profile of $S_1$ TSS and $S_2$ surface states for calculated for 5~nm Bi$_{0.85}$Sb$_{0.15}$ (12 BLs).} Values of the density of states (grey) and $\sigma_y$-polarized density of states (red) of $S_1$ TSS in Bi$_{0.85}$Sb$_{0.15}$(5nm) as a function of the bilayer index $n$ near (a) $k_x=0.0705$~\ang$^{-1}$, (b) $k_x=0.105$~\ang$^{-1}$~and (c) $k_x=0.322$~\ang$^{-1}$.}
    \label{figS_evanescence}
  \end{center}
\end{figure}

Fig.~\ref{figS_evanescence} displays the typical evanescence of both the $S_1$ (TSS) and $S_2$ surface states in depth of the Bi$_{0.85}$Sb$_{0.15}$ layer. Figs.~\ref{figS_evanescence}\reb{a,b,c} displays both DOS (red) and spin-resolved DOS (red) projected on each BL ($n=1-12$) calculated at the Fermi level ($\varepsilon=0$) near the $\overline{\Gamma}$ point for $k_\parallel=0.0705$~\AA$^{-1}$  (Fig~\ref{figS_evanescence}\reb{a} for $S_1$), 0.105~\AA$^{-1}$ (Fig~\ref{figS_evanescence}\reb{b} for $S_2$) and 0.322~\AA$^{-1}$ (Fig~\ref{figS_evanescence}\reb{c} for $S_2$ hole pocket) respectively. One note the short decrease of both DOS and spin-resolved DOS at both interfaces of the order of 2~BLs ($\lambda_{evan.}\simeq 0.5$~nm) in the exact range of the results given by Ishida for pure Bi~\cite{ishida2016}. Note also that the spin-DOS profile is exactly asymmetric from the symmetry point at the center of the layer whereas the DOS profile remains symmetric.

\subsection{TB calculation of the inverse Rashba-Edelstein tensor for Bi$_{1-x}$Sb$_x$.}

\begin{figure}[!htp]
      \includegraphics[width=\textwidth]{figS_IREEthick.png}
       \caption{\textbf{TB model and IREE conversion efficiency.} (a) Density of states (DOS) per atom as a function of the energy for Bi$_{0.79}$Sb$_{0.21}$(5nm)(12 BL). (b) BL-selected DOS in the bandgap region mainly involving the two states $S_1$ and $S_2$. (c) IREE length $\Lambda_{xy}^{\text{IREE}}$ calculated in the [-2;2] energy window emphasizing on the $S_1$ and $S_2$ contributions in the bandgap. (d) BL-selected contribution to the IREE length $\Lambda_{xy}^{\text{IREE}}$ from the 1st to the 12th bilayer of Bi$_{1-x}$Sb$_x$(5nm) for $x=0.21$. (e) IREE tensor $\kappa_{xy}$ as a function of the Bi$_{0.79}$Sb$_{0.21}$ thickness (2.5, 5, 15 nm) given in units of the spin Hall conductivity (\Ohm$^{-1}$.cm$^{-1}$). (f) Thickness-renormalized $\kappa_{xy}$ given in units of $\Lambda_{xy}^{\text{IREE}}$ (nm).}
    \label{figS_IREEthick}
\end{figure}

The IREE inverse Edelstein length $\Lambda_{xy}^{\text{IREE}}$ for Bi$_{1-x}$Sb$_x$ has been calculated within the TB framework. Fig.~\ref{figS_IREEthick}\reb{a} reports on typical results obtained for Bi$_{0.79}$Sb$_{0.21}$(5nm)(12 BLs). Fig.~\ref{figS_IREEthick}\reb{a} first displays the density of states (DOS) per atom $\mathcal{N}_\text{DOS}(\varepsilon)$ \textit{vs.} the binding/excitation in the -20 to 10~eV window. We clearly observe the presence of the valence (BVB) and conduction bulk bands (BCB) separated by the bandgap marked as dotted red lines where the two surface states $S_{1}$ and $S_2$ lie. We focus here on this region of interest in Fig.~\ref{figS_IREEthick}\reb{b} displaying the corresponding DOS per atom for each BL (from the 1st to the 12th) from -0.5 eV to 0.4 eV. One observes that the DOS is not zero in the gap (-0.2 eV - 0.2 eV) owing
to the presence of the 4 different SS ($S_1$ and $S_2$ TSS at the
respective two top and bottom surfaces). By calculating the BL-selective DOS, we can observe an enhancement of the
DOS corresponding to the two outwards BLs, labelled by
$n=1$ and $n=12$, corresponding to the two interfaces. It also proves
the strong localization/evanescence of the SS in the depth of the layers as calculated by Ishida by DFT in the
case of pure Bi~\cite{ishida2016}. A significant contribution originates from the first and 12th BL which presents a larger density of states owing to the presence of the two surface states $S_1$ TSS and $S_2$. 

The IREE length $\Lambda_{xy}^{\text{IREE}}$ is calculated by considering an intraband (spin-) relaxation time of $\tau_s=10$~fs integrating its contribution from the top BL over the mid BiSb thickness where the top evanescent states disappear. We observe in Fig.~\ref{figS_IREEthick}\reb{d} that the most important amplitude to the IREE length originates from DOS generated by $S_2$ and $S_1$ giving rise to a value of about 0.4~nm at the Fermi level ($\tau_s=10$~fs). The strong contribution of the IREE response in the bulk valence bands near -1 eV comes from a important spin-texture of surface states or bulk valence bands, which are however hard to probe with our set-up configuration. We present in Fig.~\ref{figS_IREEthick}\reb{e} the IREE tensor in the bandgap for different BiSb thickness (respectively 2.5, 5 and 15 nm). We observe that $\Lambda_{xy}$  follows the same energy dependence and increases on reducing the BiSb thickness. To extract the interfacial IREE figure of merit, we have plotted $\kappa_{xy}$ integrated over the volume (Fig.~\ref{figS_IREEthick}\reb{e}). We recover an almost-independent interconversion profile with the BiSb thickness, which is a strong element for interfacially-mediated interconversion. The extrema of this IREE tensor coincides with the $S_2$ and $S_1$ TSS.

\end{sloppypar}

\bibliographystyle{ieeetr}
\bibliography{biblio}